\documentclass{article}

\usepackage{microtype}
\usepackage{graphicx}
\usepackage{booktabs} 
\usepackage{hyperref}

\usepackage{multirow}

\usepackage[accepted]{icml2024}

\usepackage{amsmath}
\usepackage{amssymb}
\usepackage{mathtools}
\usepackage{amsthm}
\usepackage{amsfonts}
\usepackage{array}

\usepackage{comment}
\usepackage{algorithm}
\usepackage{algorithmic}
\usepackage{cancel}
\usepackage{enumerate}
\usepackage{makecell}
\usepackage[caption=false,font=normalsize,labelfont=sf,textfont=sf]{subfig}
\usepackage[export]{adjustbox}
\usepackage[T1]{fontenc}
\usepackage[capitalize,noabbrev]{cleveref}

\theoremstyle{plain}

\theoremstyle{definition}

\theoremstyle{remark}

\usepackage[textsize=tiny]{todonotes}

\icmltitlerunning{Graph Neural Network Explanations are Fragile}

\begin{document}

\twocolumn[
\icmltitle{Graph Neural Network Explanations are Fragile}

\icmlsetsymbol{equal}{*}

\begin{icmlauthorlist}
\icmlauthor{Jiate Li}{NCU,IIT}
\icmlauthor{Meng Pang}{NCU}
\icmlauthor{Yun Dong}{MSOE}
\icmlauthor{Jinyuan Jia}{PSU}
\icmlauthor{Binghui Wang}{IIT,intern}
\end{icmlauthorlist}

\icmlaffiliation{intern}{Jiate did this research as an intern in Wang's lab} 
\icmlaffiliation{NCU}{Nanchang University, China}
\icmlaffiliation{MSOE}{Milwaukee School of Engineering, USA}
\icmlaffiliation{PSU}{The Pennsylvania State University, USA}
\icmlaffiliation{IIT}{Illinois Institute of Technology, USA}

\icmlcorrespondingauthor{Meng Pang}{mengpang@ncu.edu.cn}
\icmlcorrespondingauthor{Binghui Wang}{bwang70@iit.edu}

\icmlkeywords{Graph Neural Networks, GNN Explainers}

\vskip 0.3in
]

\printAffiliationsAndNotice{}  

\begin{abstract}

Explainable Graph Neural Network (GNN) has emerged recently to foster the trust of using GNNs. Existing GNN explainers are developed from various perspectives to enhance the explanation performance. 
We take the first step to study
GNN explainers under adversarial attack---
We found that an adversary slightly perturbing graph structure 
can ensure
GNN model makes correct predictions, but the GNN explainer yields a drastically different explanation on the perturbed graph.  
Specifically, we first formulate the attack problem under a practical threat model (i.e., the adversary has limited knowledge about the GNN explainer and a restricted perturbation budget).
We then design two methods (i.e., one is loss-based and the other is deduction-based) to realize the attack.  
We evaluate our attacks on various GNN explainers 
and the results show these explainers are fragile.\footnote{Code is at: {https://github.com/JetRichardLee/Attack-XGNN}} 
\end{abstract}
\section{Introduction}
Graph Neural Network (GNN)~\citep{scarselli2008graph,kipf2017semi,hamilton2017inductive,velickovic2017graph,xu2018how} is
the mainstream paradigm for learning graph data: it takes a graph as input and learns node or graph representations to capture the relation among nodes or graph structural information. GNNs have achieved state-of-the-art performance in  graph-related tasks such as node/graph classification and link prediction
\cite{wu2020comprehensive}.

\begin{figure}[!t]
\vspace{-4mm}
    \centering
    \captionsetup[subfloat]{labelsep=none, format=plain, labelformat=empty}

    \subfloat[(a)]{
        \includegraphics[width=0.5\linewidth]{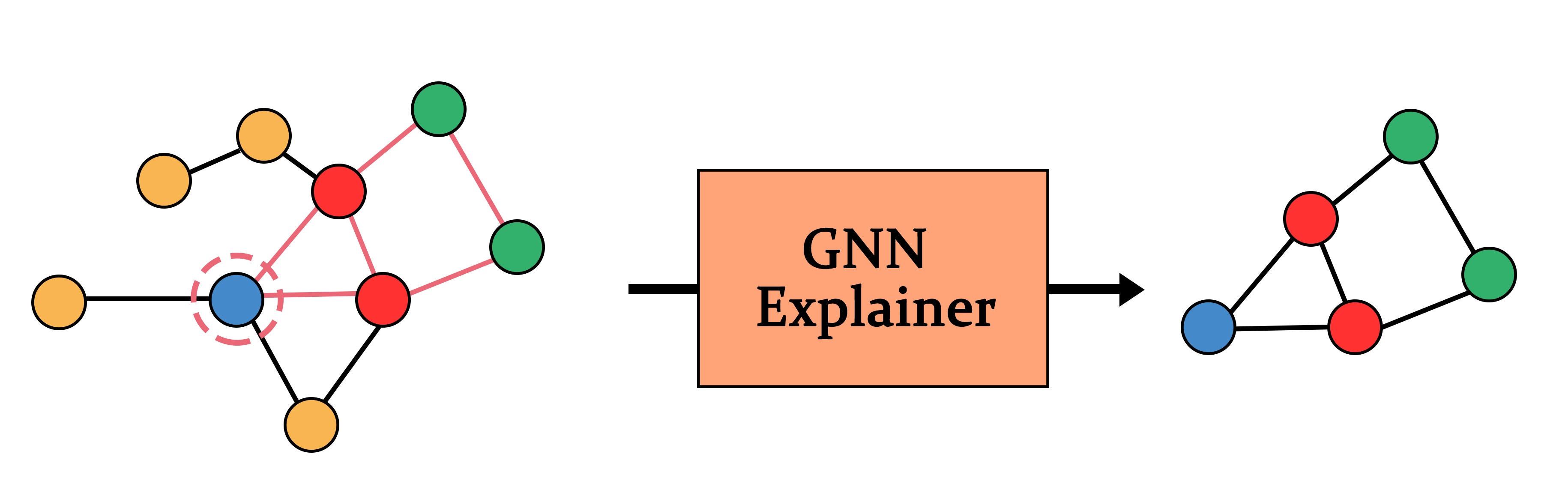}
    }\hfil
    \subfloat[(b)]{
        \includegraphics[width=0.45\linewidth]{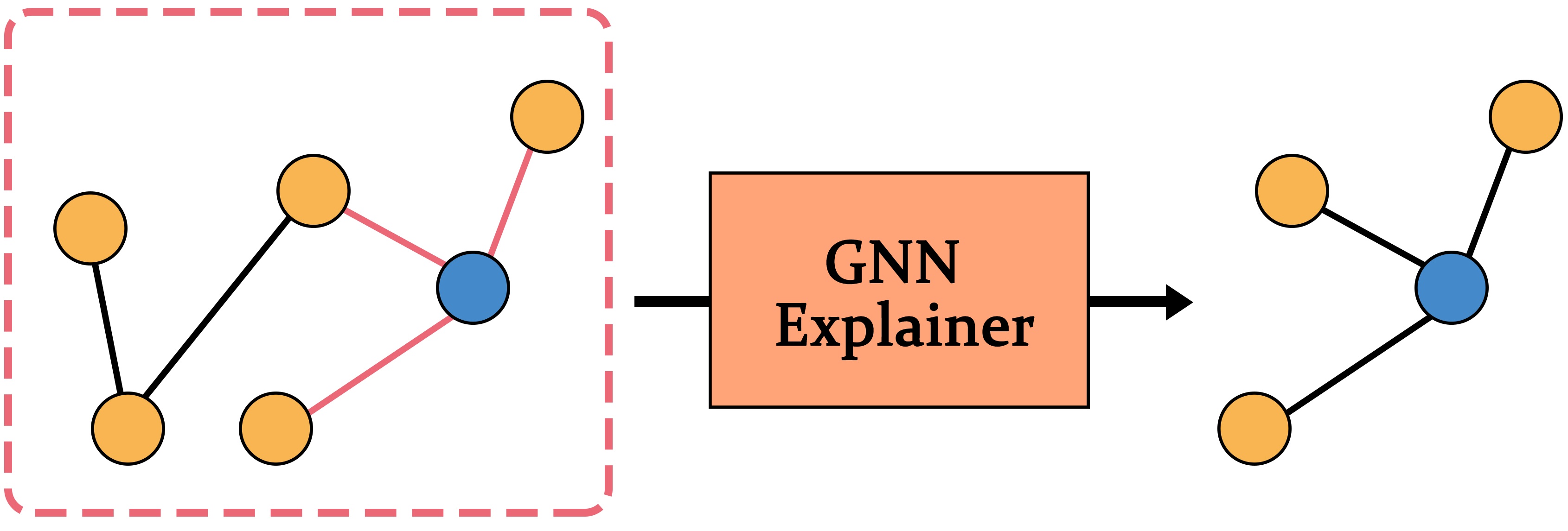}
    }
    \vspace{-2mm}
    \caption{GNN explanation for (a) node classification and (b) graph classification---It 
    identifies the subgraph that ensures the best prediction for the target node and target graph, respectively. 
    }
    \label{fig:Explanation}
   \vspace{-4mm}
\end{figure}

Explainable GNN has emerged in recent years and 
has wide applications including molecular property prediction~\cite{wu2023chemistry}, disease diagnosis~\cite{pfeifer2022gnn}, drug analysis~\cite{yang2022mgraphdta}, and fake news spreader prediction~\cite{rath2021scarlet}.
Concretely, given a graph and a predicted (node/graph) label by a GNN model, explainable GNNs 
aim to 
determine the subgraph (include edges and the connected nodes)  
 from the graph that ensures the best predictability about the label (see Figure \ref{fig:Explanation} an example). This subgraph is also called \emph{explanatory subgraph}.
 To achieve the goal, various GNN explainers from different perspectives~\cite{GNNEx19,DBLP:journals/corr/abs-2011-04573/PGExplainer,DBLP:journals/corr/abs-2102-05152/subgraphX,zhang2022gstarx,wang2023/gnninterpreter} have been proposed. 
Among them, perturbation-based explainers~\cite{GNNEx19,DBLP:journals/corr/abs-2011-04573/PGExplainer,schlichtkrull2021interpreting,funke2022z,wang2021towards,DBLP:journals/corr/abs-2102-05152/subgraphX,zhang2022gstarx} are widely studied and show promising explanation performance (more details are in Section~\ref{sec:relatedxgnn}). 

In this paper, instead of enhancing the explanation performance, we focus on understanding the robustness of GNN explainers when facing adversaries. 
Particularly, we study the research problem: \emph{Given a graph, a well-trained GNN model, and a GNN explainer, can an adversary manipulate  the graph structure such that the GNN predictions are still accurate, but the explanation result of the GNN explainer is drastically changed?}\footnote{Many existing studies propose to attack GNNs for the   classification purpose.
Their goal is different from ours and more detailed discussions about the differences are shown in Appendix~\ref{supp:discussion}.}
We emphasize this problem has serious security implications in real  applications. 
Lets take GNN-based malicious user detection in social networks 
as an instance. Assume a Facebook user is predicted as malicious by a GNN. When applying a GNN explainer, a reasonable explanation would be that the user has connected with other users that also exhibit malicious activities. Now assume an adversary has carefully added connections between the user and certain normal users.
If the user is still predicted as malicious, the explainer may wrongly interpret this is because the user has connections with those normal users, and suggest flagging them as malicious users.

We take the first step on attacking the  
most widely-studied perturbation-based 
GNN explainers with graph structure perturbations (i.e., add new edges to or/and remove existing edges from the graph).
We first characterize the threat model of the attack under three aspects: \emph{1) Attack goal}---ensure the difference between the explanation result with our attack and that without attack be as large as possible; \emph{2) Attack knowledge}---know the explanation loss (since we want to attack the specific explainer), while unknown to the internal model and training details of the explainer; 
\emph{3) Attack constraint}---have limited perturbation budget, maintain the graph structure information, and ensure correct predictions (like no attacks).   
We then formalize our attack based on the threat model. However, the attack problem is NP-hard and thus challenging to be solved. 
To address it, we propose to relax the problem and find approximate solutions, leading to our two attack design---one is loss-based and the other is deduction-based. Specifically, our two attacks are inspired by an observation: a feasible attack should be able to efficiently score
edges based on their importance and then identify the most important (added and deleted) edges to perturb the graph structure. 

{\bf Loss-based attack:} Perturbation-based GNN explainers are aiming to minimize an explanation loss (see Equation \ref{eqn:xgnnloss}).  
 This attack then uses the loss change induced by modifying an edge to indicate the edge importance. 
Particularly, an existing edge  largely increasing the loss  when it is deleted, or a new edge  largely increasing the loss when it is added, are deemed important. 
This attack then designs a new loss to capture this property. When optimized, it can uncover important edges to be perturbed. 

{\bf Deduction-based attack:} A drawback of the loss-based attack is it lacks a direct connection to the formulated attack objective. Our deduction-based attack aims to mitigate this drawback. Particularly, the key idea is to simulate the dynamic learning process of the perturbation-based GNN explainers. Then the attack objective can be rewritten in the form of a loss that has a close relationship with the loss used by the GNN explainer. It then optimizes this new loss and identifies important edges to be perturbed.

We systematically evaluate our attacks on multiple graph datasets, GNN tasks, and perturbation-based GNN explainers. Our experimental results show that existing GNN explainers are fragile.  For instance, when perturbing only 2 edges, the explanatory edges can be 70\% different from those without the attack. 
Our results also show the generated perturbations transfer well to  attacking other types of GNN explainers, thus demonstrating the generality. 
For instance, only 2 perturbed edges generated by our attacks ensure {50\%} explanatory edges in a state-of-the-art surrogate-based GNN explainer be changed.    
Our key contributions are as below:
\vspace{-2mm}
\begin{itemize}
\item To our best knowledge, this is the first work to comprehensively understand the robustness of perturbation-based GNN explainers against graph perturbations. 
\item We design two attacks on perturbation-based GNN explainers. 
Our attacks are practical (i.e., limited knowledge on the GNN explainer), stealthy (i.e., small perturbation and maintain graph structure) , and faithful (i.e., correct GNN predictions). 
\item We evaluate our attacks on multiple graph datasets, GNN tasks, and diverse types of GNN explainers.  
\end{itemize}

\section{Related Work}
\label{sec:relatedxgnn}

{\bf Explainable Graph Neural Networks:} Generally, 
GNN explainers  can be classified as \emph{instance-level based} and \emph{model-level based}. Instance-level based methods provide input instance-dependent explanation and identify the important part of a graph (e.g., edges, nodes, and features) for predicting the input instance. In contrast, model-level based methods (e.g., XGNN~\cite{DBLP:journals/corr/abs-2006-02587/XGNN} and GNNInterpreter~\cite{wang2023/gnninterpreter}) do not consider specific input instances, but generate graph patterns to explain a class of instances. 
In the paper, we focus on instance-level explainers 
as they are more widely studied. 

Instance-level GNN explainers can be further classified into five categories: \textit{decomposition-based}, \textit{gradient-based},  \textit{surrogate-based}, \textit{generation-based}, and \textit{perturbation-based}. Details of these categories are seen in Appendix~\ref{supp:relatedwork}. 
We primarily focus on {perturbation-based methods}, as their explanation results are more accurate.

{\bf Attacking Neural Network Explanations:}
These studies ~\cite{interpretationFragile,explanationsManipulated,DBLP:journals/corr/abs-2106-02666/Manipulated,foolingOnNNI} have explored the robustness of explanations, primarily on image models, against adversarial attacks. 
Particularly, their goal is  
introducing subtle perturbations to input images that do not alter image predictions, but can drastically change the explanations.
As we will see, attacking explainable GNN methods is much more challenging in that the attack has more constraints and the attack problem is essentially NP-hard.

{\bf Attacking Graph Neural Networks:}
Existing attacks on GNNs all focus on misleading classification models  and they can be categorized as \emph{test-time evasion attacks}~\cite{dai2018adversarial,zugner2018adversarial,xu2019topology,wu2019adversarial,ma2020towards,mu2021hard,wang2022bandits,wang2023turning,wang2024efficient} and \emph{training-time poisoning attacks}~\cite{xu2019topology,zugner2019adversarial,wang2019attacking,wang2023turning}. Take GNNs for node classification as an instance. In evasion attacks, given a trained GNN model and a (clean) graph, an attacker carefully perturbs the graph structure
such that the GNN model misclassifies as many testing nodes as possible in the perturbed graph. In poisoning attacks, given a GNN algorithm and a graph, an attacker carefully perturbs the graph structure in the training phase, such that the learnt GNN model misclassifies as many testing nodes as possible in the testing phase. 

A closely relevant work to ours is the 
GEAttack 
\cite{fan2022jointlyAttackExplanation}. 
However, the threat model and attacker goal are different from ours. 
First, GEAttack has white-box access to the GNN explainer; Second, GEAttack also aims to alter GNN \emph{predictions}---it  
{adds} new edges to a graph such that 
the GNN classifier produces \emph{wrong} predictions for nodes in the perturbed graph and the added edges are \emph{within} the explanatory subgraph outputted by the GNN explainer. 

\section{Background}

\subsection{Perturbation-based GNN Explainers}
Given a graph $G=(V,E)$ with node set $V$, edge set ${E}$, 
a label $y$ (on a node $v \in V$ or the entire graph $G$); and a well-trained GNN model $f$ with an accurate prediction, i.e.,  $f(G) = y$.
The objective of GNN explainers is 
identifying a subgraph
\footnote{
Some GNN explainers (e.g., \citet{GNNEx19}) also interpret 
node features, which are 
not as effective as graph structure.} $G_S = (V_S, E_S) \subset G$
that ensure $f$'s best predictability for the label $y$, i.e., $\max_{G_S} Pr(f(G_S)=y)$. 
Note that when the important edges $E_S$ are determined, the connected nodes $V_S$ are determined accordingly. For simplicity, many GNN explainers hence focus on identifying  $E_S$ which is called the \emph{explanatory edges}. 

In this paper, we consider the widely-studied perturbation-based GNN explainers and briefly review it as below:

\textbf{1.} An edge mask $M \in [0,1]^{|{E}|}$ is defined and initialized deterministically or stochastically. 
This induces a masked graph $G_M$ with the masked edges defined as $E \otimes M $, where $\otimes$ means the element-wise product.  A mask value $M_e$ indicates the important score of an edge $e \in E$. For instance, $M_e=1$ means $e$ is extremely important, while $M_e=0$ means $e$ is extremely unimportant for prediction.

\textbf{2.} A GNN explainer-dependent objective function $\mathcal{L}$:\footnote{We omit $E$ in the loss
$\mathcal{L}$ and its meaning is clear from context.}
{\vspace{-1mm}
\begin{equation}
\label{eqn:xgnnloss}
\mathcal{L}(M) = l(f(G_M),y) + \mathcal{C}(M),
\end{equation}
}%
where $l$ denotes the prediction loss  with respect to the masked graph $G_M$; 
and  $\mathcal{C}$ is a constraint function on the mask. The prediction loss and constraint functions vary among different explainers.  Table~\ref{table:explainer} shows these functions in representative perturbation-based GNN explainers. 

\textbf{3.} The edge mask is optimized to decide the explanatory edges.  
First, the GNN explainer learns the mask $M$ by minimizing the loss $\mathcal{L}(M)$ via gradient descent: $M = M - r \cdot {\partial \mathcal{L}(M)}/{\partial M}$, where $r$ is the learning rate. 
Then, the edges with the $k$ highest  values in the learnt mask $M$ are selected as the explanatory edges ${E}_{S}$:
{
\begin{equation}\label{eo}
{E}_{S} = E.\text{top}_k(M).
\end{equation}
}

\subsection{Power-Law Likelihood Ratio Test} 
\label{sec:ratiotest}
This test is used to measure the similarity between two graphs. 
Suppose we have a graph $G$ and a generated graph $G_A$. 
To ensure 
$G_A$ possesses similar structural properties as
$G$, we employ a two-sample test for power-law distributions, which mainly evaluates the likelihood between the degree distributions of both $G$ and $G_A$. 

Given a power-law distribution $p(x) \propto x^{-\alpha}$, the initial step involves estimating the scaling parameter $\alpha$. Although there exists no precise solution for discrete data, such as the degree distribution \cite{bessi2015samples,zugner2018adversarial} proposed an approximate expression for $G$ (or $G_A$) as:
{\vspace{-1mm}
\begin{equation}
\alpha_{G} = 1 + |\mathcal{D}_{G}| \cdot [\Sigma_{d_{i} \in \mathcal{D}_{G}} \log \left({d_{i}}/{(d_{\text{min}} - {1}/{2})} \right)]^{-1}
\end{equation}
}%
Here, $d_{\text{min}}$ is the minimal degree of nodes in a test set. 
$\mathcal{D}_{G} = \{d_{v}^{G} \mid v \in V, d_{v}^{G} \geq d_{\text{min}}\}$ is a multi-set containing node degrees. Using the obtained scaling parameter $\alpha_{G}$ for sample $\mathcal{D}_{G}$, its log-likelihood can be assessed by:
{\vspace{-1mm}
{\begin{align}
l(\mathcal{D}_{G}) & = |\mathcal{D}_{G}| \cdot \log \alpha_{G} + |\mathcal{D}_{G}| \cdot \alpha_{G} \cdot \log \alpha_{G} \nonumber \\ 
& - (\alpha_{G} + 1) \sum_{d_{i} \in \mathcal{D}_{G}} \log d_{i}
\end{align}}
\vspace{-5mm}}

Similarly, we can  compute the log-likelihood for $\mathcal{D}_{G_A}$ and for 
$\mathcal{D}_{G} \cup \mathcal{D}_{G_A}$.  
The ratio test statistic is then given by:
\begin{equation}
\Lambda(G, G_A) = -2 \cdot l(\mathcal{D}_{G} \cup \mathcal{D}_{G_A}) + 2 \cdot l(\mathcal{D}_{G}) + 2 \cdot l(\mathcal{D}_{G_A})
\end{equation}
This statistic adheres to a $\chi^{2}$ distribution with a single degree of freedom. For instance, in this paper, we require that the generated graph $G_A$ exhibits more than 99\% structural similarity with $G$. Then, we only accept $G_A$ when: $\Lambda(G, G_A) < \tau \approx 0.000157$. 

\section{Our Attack Design}
\label{sec:reverse}

In this section, we present our attack method in detail. 
We first define the threat model that characterizes the attacker's  goal, knowledge, and constraints (e.g., stealthiness, faithfulness, practicability).   
We then formalize our attack by integrating the threat model, and propose two attack methods to solve the attack problem. 

\subsection{Attack Formulation}
\label{sec:formulation}

{\bf Threat model:} 
We first characterize our threat model\footnote{{Our threat model is suitable for practical scenarios where GNN explainers are as-a-service (e.g., they are deployed as an API to provide visualized explanations with users' input graphs). For instance, Drug Explorer \cite{wang2022extending} is a recent explainable GNN tool for drug repurposing (reuse existing drugs for new diseases), where users input a drug graph and the tool outputs the visualized explanation result. 
}}.

\emph{Attack Goal:} Given an explanation instance $\{G, y, f,  {E}_{S}\}$, an attack introduces edge perturbations (via injecting new edges or deleting the existing edges) to the graph $G$, which produces 
a  modified edge set $\tilde{E}$ (and a modified graph $\tilde{G}$) 
with a new mask $\tilde{M}$ after the attack. As the edge status of any pair of nodes in the graph can be perturbed and hence the attack mask $\tilde{M} \in [0,1]^{|V|^2}$.  
GNN explainers can then yield a new explanation $\tilde{E}_{S}$ based on $\tilde{M}$:
{
\begin{align}\label{ep}
\tilde{M} &= \underset{\tilde{M}}{{\arg\min}} \, \mathcal{L}(\tilde{M}) =
l(f({\tilde{G}_{\tilde{M}}}),y) + \mathcal{C}(\tilde{M}),\\
\tilde{E}_{S} &= \tilde{E}.\text{top}_k(\tilde{M}), \label{exp_attack} 
\end{align}
}%
The attack goal is to ensure the difference between $\tilde{E}_{S}$ after the attack and $E_S$ without attack be as large as possible.

\emph{Attack Knowledge:}
In real-world applications, the attacker often has limited knowledge about the 
GNN explainer. We consider this scenario
in this paper. Specifically, we assume the attacker only knows the explanation loss $\mathcal{L}$ (as s/he wants to attack a specific explainer) 
and the explanatory edges $E_S$ (not the mask values $M_{E_S}$) outputted by the explainer.
 The attacker does not know the internal model parameters or  architecture, nor the  training details of the explainer. 

\emph{Attack Constraint:} To ensure our attack be realistic,
we consider the following constraints on the attack:
\begin{enumerate}[1]
\vspace{-2mm}
\item The explanatory edges ${E}_{S}$ 
on the clean graph $G=(V,E)$ should be 
 kept in 
 the perturbed graph $\tilde{G}=(V,\tilde{E})$. Otherwise, there is a trivial solution where the attacker can simply remove 
 ${E}_{S}$ to fool the GNN explainer.

\vspace{-1mm}
\item The number of perturbed edges 
should not exceed a perturbation budget $\xi$. This is to enforce that the attacker has constrained resources and make the attack practical. 

\vspace{-1mm}
\item 
$\tilde{G}$ and $G$ should be structurally similar. This ensures the edge perturbations are stealthy, i.e., hard to be detected via checking the graph structure. Here, we 
follow the power-law likelihood ratio test in Section~\ref{sec:ratiotest} to measure graph structure similarity.  
\vspace{-1mm}
\item The GNN model $f$ still makes accurate predictions on the perturbed graph. This ensures the model faithfulness. 
\vspace{-2mm}
\end{enumerate}

\noindent {\bf Problem formulation:}
Based on the above threat model, 
we formalize  our attack as below:
{
\begin{align}
&\underset{\tilde{E}}{{\arg\max}} \, |{E}_{S}- \tilde{E}_{S}\cap E_S| \label{problem} \\
s.t.\quad  & {E}_{S}\subseteq \tilde{E} \label{eqn:c1} \\ 
& |{E}\cup \tilde{E}-{E}\cap\tilde{E}| \leq \xi  \label{eqn:c2} \\
& \Lambda(G,\tilde{G}) < \tau  \label{eqn:c3} \\
& f(\tilde{G}) = y \label{eqn:c4}
\end{align}
}

However, the objective function of Equation~\eqref{problem} involves the set operation, which makes it hard to be optimized.
Next, we propose to relax the objective function defined on the set to be that defined on numeral value.  
Let's first consider the ideal case, where the mask values are binary.
We observe that the sum of the attack mask $\tilde{M}$'s values on the explanatory edges $E_S$ (without attack), denoted as an 
$\ell_1$-norm  $\|\tilde{M}_{E_S}\|_{1}$, equals to the set size of $|\tilde{E}_{S}\cap E_S|$. Then we could rewrite 
Equation~\eqref{problem} as:
{
\begin{equation*}
\begin{aligned}
   |E_S-\tilde{E}_{S}\cap E_S| &= |E_S|-|\tilde{E}_{S}\cap E_S|
   =k - \|\tilde{M}_{E_S}\|_{1}.
\end{aligned}
\end{equation*}
}%
As $k$ is a constant, Equation~\eqref{problem} 
could be reformulated as:
{\vspace{-1mm}
\begin{equation}
\label{eq:problem2}
\underset{\tilde{E}}{\text{argmin}} \quad \|\tilde{M}_{E_S}\|_{1}
\end{equation}
}%
This means the attack objective is now to learn an attack mask $\tilde{M}$ on the perturbed graph $\tilde{G}$ that maximally decreases the mask values on $E_S$. 
However, due to the attack constrains on the sets, this attack optimization problem is combinatorial, which is NP-hard (a detailed proof is provided in Appendix~\ref{app:proveNP}).
This implies determining the optimal edge perturbations within polynomial time is impossible. 
To address it, we propose our attack that finds an approximate  solution to solve this NP-hard problem.

\begin{figure}[t]
\centering
\includegraphics[width=\linewidth]{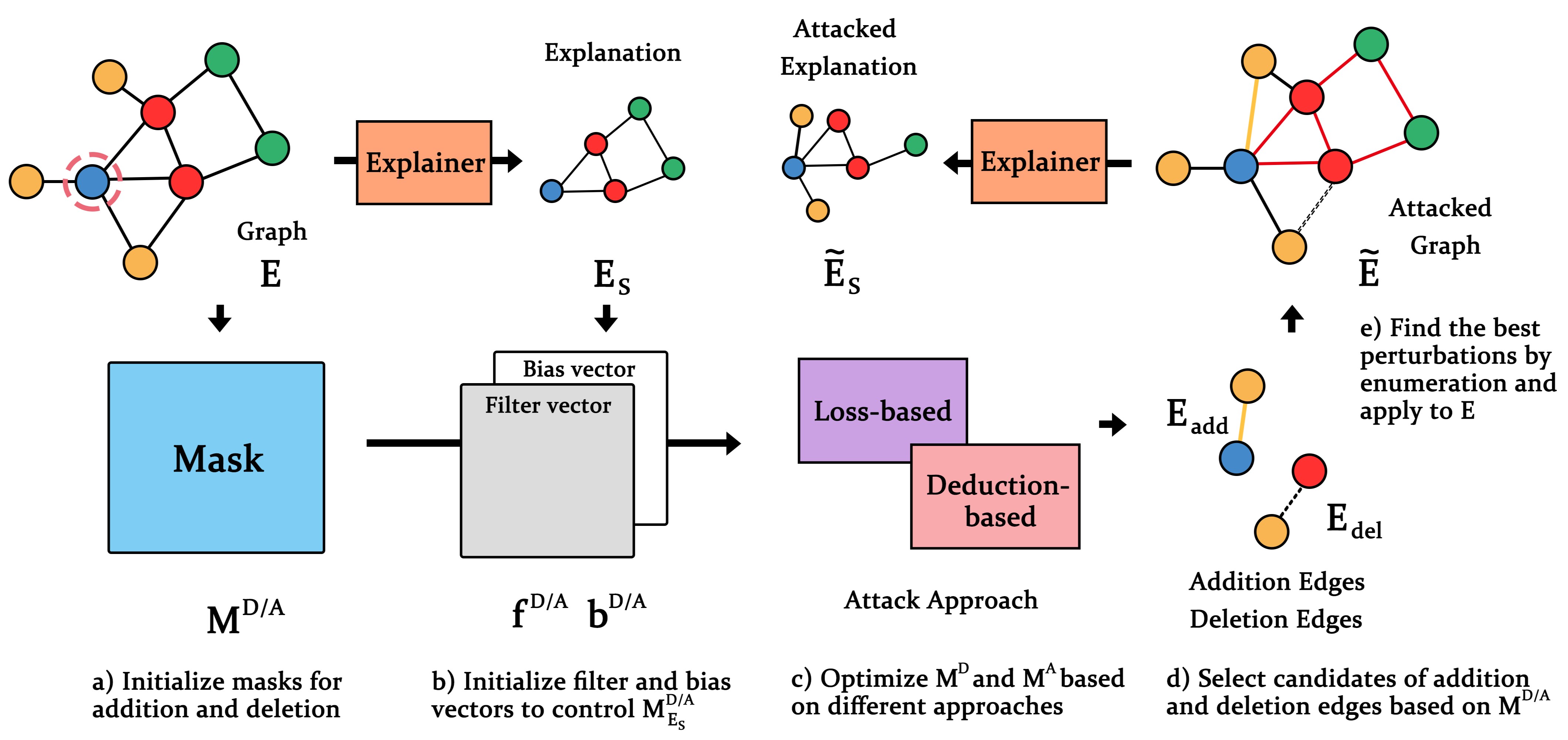}
\vspace{-4mm}
\caption{Overview of our attacks. Take node classification for instance: we find perturbations that maximize the difference between $E_{S}$ and $\tilde{E}_{S}$ while satisfying the attack constraints.}\label{fig:ReverseLearning}
\vspace{-2mm}
\end{figure}

\subsection{Attack Methodology}
\label{sec:attack}

Our attack is inspired by the observation: A majority of the edges are not important for prediction/explanation,  hence 
a feasible attack should be able to efficiently \emph{score  edges based on their  importance 
and then identify the most important edges for addition and deletion.} 
Next we propose our two attack methods based on this motivation. 

\subsubsection{Loss-based Attack} \label{greedy}

This attack method uses the loss change induced by modifying an edge to indicate the edge importance. Note that perturbation-based GNN explainers aim to minimize an explanation loss defined in Equation \ref{eqn:xgnnloss}. 
Hence, we deem an edge as important if it is an existing edge and largely decreases the  loss  when deleted, or it is a new edge and largely increases the loss when it is added. Deleting these existing edges or/and adding these new edges can significantly affect the explanatory edges  $E_{S}$. 
Hence, this attack aims to uncover the edges that largely change the loss.

\noindent {\bf Loss on $\tilde{M}^D$ and $\tilde{M}^A$:} 
Obviously, added edges are restricted to the complementary edges in $G$, denoted as  $E^C = \{(i, j): \forall i, j \in V, (i,j) \notin E, i \neq j\}$, while deleted edges are from the existing edges $E$ in $G$. 
Due to the {\bf Attack Constraint 1)}, we do not allow deleting the explanatory edges $E_S$ and denote the edge deletion set as $E^D = E - E_S$.  
We also denote $E^A$ as $ E^A =E^C \cup E_{S}$ for description simplicity. 
With it, we have $|E^A| + |E^D| = |\tilde{M}|$.  

Now we split the attack mask $\tilde{M}$ into an edge deletion mask  $\tilde{M}^{D} \in [0,1]^{|E|}$ on $E$ and an edge addition mask $\tilde{M}^{A} \in [0,1]^{|E^A|}$ on $E^A$. 
We then require $\tilde{M}^D$ and $\tilde{M}^A$  trainable while not learning the mask value on $E_S$ (e.g., set a constraint $\tilde{M}^{D}_{E_{S}} = \gamma$ and $\tilde{M}^{A}_{E_{S}} = \gamma$) during training.    
Motivated by the perturbation-based GNN explanation, 
we design below two losses on $\tilde{M}^D$ and $\tilde{M}^A$ respectively: 
{
\begin{align}
\label{eqn:greedy}
& \min_{\tilde{M}^D} \mathcal{L}(\tilde{M}^D)|_{\tilde{M}^D_{{E_S}}=\gamma}, \quad \max_{\tilde{M}^A} \mathcal{L}(\tilde{M}^A)|_{\tilde{M}^A_{{E_S}}=\gamma} 
\end{align}
}%
where $\mathcal{L}(\tilde{M}^D)|_{\tilde{M}^D_{E_S}=\gamma}$ means 
$\mathcal{L}(\tilde{M}^D)$ defined on 
$\tilde{M}^D$ with values on $E_S$ set to be $\gamma$, and similar to $\mathcal{L}(\tilde{M}^A)|_{\tilde{M}^A_{E_S}=\gamma}$. 

To make the above losses trainable, we introduce four vector variables: ${\bf f}^{D}, {\bf b}^{D} \in \{0,1\}^{|E|}$, and ${\bf f}^{A}, {\bf b}^{A} \in \{0,1\}^{|E^A|}$, where ${\bf f}^{D}$ and ${\bf f}^{A}$ are called filter vectors, and ${\bf b}^{D}$ and ${\bf b}^{A}$ are called bias vectors for $\tilde{M}^D$ and $\tilde{M}^A$, respectively. And we set the values as below: 
{
\begin{equation}
\begin{aligned}
& {\bf f}^D_{e} = 0,  \, {\bf b}^D_{e} = 1,  \text{ if } e \in E_S; \, {\bf f}^D_{e} = 1,  \, {\bf b}^D_{e} = 0, \text{ if } e \in E^D.
\\
& {\bf f}^A_{e} = 0,  \, {\bf b}^A_{e} = 1,  \text{ if } e \in E_S; 
{\bf f}^A_{e} = 1,  \, {\bf b}^A_{e} = 0, \text{ if } e \in E^C. \label{fb_MA}
\end{aligned}
\end{equation}
}%
Then we have: 
{\small
\vspace{-1mm}
\begin{align}
& \min_{\tilde{M}^D} \mathcal{L}(\tilde{M}^D)|_{\tilde{M}^D_{{E_S}}=\gamma} = \mathcal{L}(\tilde{M}^D \otimes {\bf f}^{D} + \gamma \cdot {\bf b}^{D}) \label{loss:MD}\\
& \max_{\tilde{M}^A} \mathcal{L}(\tilde{M}^A)|_{\tilde{M}^A_{{E_S}}=\gamma} = \mathcal{L}(\tilde{M}^A \otimes {\bf f}^{A} + \gamma \cdot {\bf b}^{A}) \label{loss:MA}
\end{align}
}%
We observe that, for any $\gamma \in [0,1]$, 
the results of $\tilde{M}^D$ and $\tilde{M}^A$ on ${E_{S}}$ are always $\gamma$.
Besides, since ${\bf f}^D_{e}=0$ and ${\bf f}^A_{e}=0$ for all $e \in E_{S}$, the mask values $\tilde{M}^{D}_{E_{S}}$ and  $\tilde{M}^{A}_{E_{S}}$  are never learnt during training. 

\noindent {\bf Candidate edges in $\tilde{M}^D$ ( $\tilde{M}^A$) for deletion (addition):} 
After learning $\tilde{M}^D$ (or $\tilde{M}^A$), 
for an edge $e \in E^{D}$ (or $e \in E^{A}$), if it has a higher mask value $\tilde{M}^{D}_{e}$ (or $\tilde{M}^{A}_{e}$), it has more positive effect on \emph{decreasing} (or \emph{increasing}) the loss w.r.t. $E_{S}$, and conversely, deleting (or adding) it could increase the loss more. We hence select the edges in $E^{D}$ (or $E^{A}$) with the highest mask values as edge deletion (or addition) candidates $E_{del}$ (or $E_{add}$).

\noindent {\bf Deciding the edges for perturbation:} To decide the best combination of deleted edges and added edges, we enumerate over different $\xi_{D}$ and $\xi_{A}$ such that $\xi_{D} + \xi_{A} = \xi$  (\emph{satisfy Attack Constraint 2}), where we select the top $\xi_{D}$ and top $\xi_{A}$ candidates in $E_{del}$ and $E_{add}$, respectively. 
A generated perturbed graph $\tilde{G}$ under a pair $(\xi_{D},\xi_{A})$ is kept, if yields a correct GNN prediction $f(\tilde{G})=y$ (\emph{satisfy Attack Constraint 3}), and passes the likelihood test $\Lambda(G,\tilde{G})$ (\emph{satisfy Attack Constraint 4}). The perturbed graph, when evaluated by a GNN explainer, leading to the largest $|E_{S}-E_{S}\cap \tilde{E}_{S}|$ will be chosen as our final attack result. 
Algorithm~\ref{alg:loss} in Appendix~\ref{supp:loss-based} shows the details of the loss-based attack. 

\noindent {\bf Drawbacks:} The loss-based attack is effective  to some extent (see our results in Section~\ref{sec:exp}). However, a key drawback is that it lacks a direct connection to the attack objective in either Equation (\ref{problem}) or Equation (\ref{eq:problem2}).  
This highlights the necessity on designing an attack method that directly considers the attack objective. Next, we propose another  deduction-based attack to mitigate the drawback.

\subsubsection{Deduction-based Attack}
Our key idea in this attack is to simulate the dynamic mask learning process by the perturbation-based GNN explainer. 
Then we evaluate the potential of deleted edges from 
$E^D$ and added edges from $E^A$ to minimize $||\tilde{M}_{E_S}^*||_1$.

\noindent {\bf Deductive reasoning:} Assume the GNN explainer 
learns the final mask ${M}^*$ 
from the initial mask ${M}^0$ on the clean graph $G$
with a loss $\mathcal{L}(M)$ defined in Equation~\ref{eqn:xgnnloss} and a constant learning rate $r$. Then, in each iteration $i \in \{1, \cdots, \infty\}$, the mask $M^{i}$ can be updated as 
{
\begin{align}
\label{eqn:gnnex_iter}
M^{i} = M^{i-1} - r \frac{\partial \mathcal{L}(M)}{\partial M} |_{M=M^{i-1}}. 
\end{align}
}%
When applied to the explanatory edges $E_S$:
{
\begin{equation*}
\label{eqn:gnnex_iter}
M^{i}_{E_S} = M^{i-1}_{E_S} - r \frac{\partial \mathcal{L}(M)}{\partial M_{E_S}} |_{M_{E_S}=M^{i-1}_{E_S}}
\end{equation*}
}%
Applying this equation 
iteratively, the final mask on $E_S$ is: 
{\small
\begin{equation*}
M^*_{E_S} = M^{0}_{E_S} - r \lim_{n\rightarrow \infty} \sum_{i=0}^{n-1} \frac{\partial \mathcal{L}(M)}{\partial M_{E_S}} |_{M_{E_S}=M^{i}_{E_S}}
\end{equation*}
}%
As a result, the $\ell_1$-norm of $M^*_{E_S}$ can be expressed as:
\begin{small}
\begin{equation*}
||M^*_{E_S}||_{1} = ||M^{0}_{E_S} - r \lim_{n\rightarrow \infty} \sum_{i=0}^{n-1} \frac{\partial \mathcal{L}(M)}{\partial M_{E_S}} |_{M_{E_S}=M^{i}_{E_S}}||_{1}
\end{equation*}
\end{small}

Given that the elements in $M^*_{E_S}$ 
are all positive (otherwise ${E_S}$ cannot be the important edges), the $\ell_1$-norm of $M^*_{E_S}$ equals to the sum of all its elements $M^*_{e}, \forall e \in E_S$:
{
\vspace{-1mm}
\small
\begin{equation}
\begin{aligned}\label{sigma}
||M^*_{E_S}||_{1} &= \sum_{e \in E_S} \left[ M^{0}_{e} - r \lim_{n\rightarrow \infty} \sum_{i=0}^{n-1} \frac{\partial \mathcal{L}(M)}{\partial M_{e}} |_{M_{e}=M^{i}_{e}} \right]\\
&=\sum_{e \in E_S} M^{0}_{e} - r\lim_{n\rightarrow \infty} \sum_{i=0}^{n-1} \sum_{e \in E_S} \frac{\partial \mathcal{L}(M)}{\partial M_{e}} |_{M_{e}=M^{i}_{e}}
\end{aligned}
\end{equation}
}%
As the real mask learning process is black-box to the attacker, we have no access to the true values of $M^{i}_{E_S}, i \in \{1, \cdots, \infty\}$. To address it, we introduce a new variable $\alpha \in [0,1]$ and its value sequence $
\alpha_{i}, \forall i \in \{1, \cdots, \infty\}$ to approximate the mask sequence $M^{i}_{E_S}, \forall i \in \{1, \cdots, \infty\}$. At each iteration $i$, we assume the mask values $M^{i}$ of important edges $E_S$ equal to (relatively large) $\alpha_{i}$:
\begin{equation}\label{approx}
M_{e} = \alpha |_{M_{e} = M^{i}_{e}, \alpha = \alpha_{i}}, \forall e \in E_S, i \in \{1, \cdots, \infty \}
\end{equation}
With this assumption, the following properties are satisfied:
{\small
\begin{equation}\label{deriva}
\frac{\partial M_{e}}{\partial \alpha} = 1 |_{M_{e} = M^{i}_{e}, \alpha = \alpha_{i}}, \forall e \in E_S, i \in \{1, \cdots, \infty \}
\end{equation}
\begin{equation}
\begin{aligned}\label{sume}
\sum_{e \in E_S} M^{0}_{e} &= \sum_{e \in E_S} \alpha_{0} = |E_S|\alpha_{0}
\end{aligned}
\end{equation}
}%
Incorporating Equations~\ref{approx}, ~\ref{deriva} and~\ref{sume} into  Equation \ref{sigma}, we rewrite Equation \ref{sigma} 
via a differential equation on $\alpha$:
{
\small
\begin{equation}
\begin{aligned}\label{alpha}
||M^*_{E_S}||_{1} &=|E_S|\alpha_{0} - r \lim_{n\rightarrow \infty} \sum_{i=0}^{n-1} \sum_{e \in E_S} \frac{\partial \mathcal{L}(M)}{\partial M_{e}} \frac{\partial M_{e}}{\partial \alpha} |_{M_{E_S}=\alpha, \alpha = \alpha_{i}}\\
&= |E_S|\alpha_{0} - r\lim_{n\rightarrow \infty} \sum_{i=0}^{n-1} \frac{\partial \mathcal{L}(M)|_{M_{E_S}=\alpha}  }{\partial \alpha}|_{\alpha = \alpha_{i}} \\
&\approx |E_S|\alpha_{0} - r\int_{\alpha_0}^{\alpha_\infty} \frac{\partial \mathcal{L}(M)|_{M_{E_S}=\alpha}  }{\partial \alpha}
\end{aligned}
\end{equation}
}%
Here, $\alpha_{0}$ represents an initial mask value and $\alpha_{\infty}$ approaches the final mask value for edges $e \in E_S$. 
Here, we set $\alpha_{0}=0$ and approximate the integral in Equation~\ref{alpha} via sampling. Specifically, we  use $N$ samples and define constants $\{\beta_{i}\}_{i=1}^N$ based on a lower bound constant $\beta \in [0,1]$: 
{
\begin{equation*}
\begin{small}
\begin{aligned}
\beta_{i} &= \frac{i-1}{N-1}\times (1-\beta)+\beta, \quad \forall i \in \{1, 2, \cdots, N \}
\end{aligned}
\end{small}
\end{equation*}
}%
Consequently, we express $\|M^*_{E_S}\|_{1}$ as below
{
\begin{small}
\begin{align*}\label{target}
||M^*_{E_S}||_{1}  &\approx -\frac{r}{N}\sum_{i=1}^N  \big(\mathcal{L}(M)|_{M_{E_S}=\beta_{i}} - \mathcal{L}(M)|_{M_{E_S}=0} \big), 
\end{align*}
\end{small}%
}%
where $\mathcal{L}(M)|_{M_{E_S}=\beta_{i} (\text{or } =0)}$ 
means 
$\mathcal{L}(M)$ on the mask $M$ with values on the explanatory edges $E_S$ set to be $\beta_{i}$ (or $0$).

As an attack, instead, we reverse the mask learning: 
{
\vspace{-1mm}
\begin{small}
\begin{align*}
||\tilde{M}^*_{E_S}||_{1}  &\approx \frac{r}{N}\sum_{i=1}^N  \big(\mathcal{L}(\tilde{M})|_{\tilde{M}_{E_S}=\beta_{i}} - \mathcal{L}(\tilde{M})|_{\tilde{M}_{E_S}=0} \big). 
\end{align*}
\end{small}
}
Then, we can follow the loss-based attack to define the loss on the edge deletion mask $\tilde{M}^D$ and edge addition mask $\tilde{M}^A$ and decide the deleted and added edges for perturbation. 

\noindent {\bf Loss on $\tilde{M}^D$ and $\tilde{M}^A$:} 
We first define a loss $\bar{\mathcal{L}}$ on 
$\tilde{M}^D \in [0,1]^{|{E}|}$ for the edge set $E$ to be minimized\footnote{W.l.o.g, we omit the constant $r/N$ for description simplicity}: 
{\small
\vspace{-1mm}
\begin{small}
\begin{align}
& \bar{\mathcal{L}}(\tilde{M}^D) = \sum_{i=1}^N \left[ \mathcal{L}(\tilde{M}^D)|_{\tilde{M}^D_{E_S}=\beta_{i}} - \mathcal{L}(\tilde{M}^D)|_{\tilde{M}^D_{E_S}={0}} \right]
\nonumber \\
& \, = \sum_{i=1}^N \left[ \mathcal{L}(\tilde{M}^D\otimes {
\bf f}^D + \beta_{i} \cdot {\bf b}^D) - \mathcal{L}(\tilde{M}^D\otimes {\bf f}^D) \right]. 
\end{align}
\end{small}
}%
Similarly, we define another loss $\bar{\mathcal{L}}$ on $\tilde{M}^A \in [0,1]^{|{E^A}|}$ for the edge set $E^A$ to be maximized: 
{\small
\vspace{-2mm}
\begin{small}
\begin{align}
& \bar{\mathcal{L}}(\tilde{M}^A) = \sum_{i=1}^N \left[ \mathcal{L}(\tilde{M}^A)|_{\tilde{M}^A_{{E_S}}=\beta_{i}} - \mathcal{L}(\tilde{M}^A)|_{\tilde{M}^A_{{E_S}}={0}} \right] \nonumber \\ 
& \, = \sum_{i=1}^N \left[ \mathcal{L}(\tilde{M}^A\otimes {\bf f}^A + \beta_{i} \cdot {\bf b}^A) - \mathcal{L}(\tilde{M}^A \otimes {\bf f}^A) \right],
\end{align}
\end{small}%
}%
where ${\bf f}^D$, ${\bf f}^A$, ${\bf b}^D$, and ${\bf b}^A$ are defined  in Equation \ref{fb_MA}.

\noindent {\bf Deciding the edges for perturbation:} 
After minimizing $\bar{\mathcal{L}}(\tilde{M}^D)$, the edges in 
$E^D$ with the highest masked values in $\tilde{M}^D$
are the ones that reduce $\|\tilde{M}_{E_S}\|_{1}$ the most  
when they are deleted from ${G}$.
After maximizing the loss $\bar{\mathcal{L}}(\tilde{M}^A)$, the edges in $E^A$ with the highest masked values in $\tilde{M}^A$ 
are the ones that reduce $\|\tilde{M}_{E_S}\|_{1}$ the most when they are added to $G$. 
We then follow the enumeration strategy proposed before 
and identify the best deleted and added edges to generate the perturbed graph as our attack.  Algorithm~\ref{alg:deduction} 
in Appendix~\ref{supp:deduction-based} details the deduction-based attack.

\section{Experiment}
\label{sec:exp}

\subsection{Experiment Setup}

{\bf Datasets:}
We evaluate GNN explanations on both node classification and graph classification tasks. For node classification, following existing works~\cite{GNNEx19,DBLP:journals/corr/abs-2011-04573/PGExplainer} we choose three synthetic datasets,
i.e., BA House, BA Community, and Tree Cycle. We also add one large real-world dataset OGBN-Products~\cite{Bhatia16}. For graph classification, we use two real-world datasets, MUTAG~\cite{kriege2012subgraph/mutagDATA} and Reddit-Binary~\cite{10.1145/2783258.2783417/RBDATA}. Example graphs of the datasets and detailed descriptions are shown in Appendix~\ref{supp:ExperimentSetting}.

{\bf Base GNN model and GNN explainers:} We use the graph convolutional network (GCN)~\cite{kipf2017semi} as the base GNN model for node and graph classification, following~\cite{GNNEx19}. The testing accuracy (all are $>80\%$) of the trained GCN on the datasets are shown in Table~\ref{table:accuracy}. 
\emph{To ensure the explanation quality, we only select the testing nodes/graphs that are correctly predicted by the GNN model for evaluation.}
We choose three well-known perturbation-based GNN explainers: GNNExplainer~\cite{GNNEx19}, PGExplainer~\cite{DBLP:journals/corr/abs-2011-04573/PGExplainer}, and GSAT
\cite{DBLP:journals/corr/abs-2201-12987/GSAT} (More details see Table~\ref{table:explainer}). 

{\bf Evaluation metric:} As different datasets use different number of explanatory edges, we evaluate the attack performance using the overlap ratio between the explanatory edges after and before the attack obtained by the explainers.  That is, ${|{E}_{S}-E_{S}\cap \tilde{E}_{S}|}/{|{E}_{S}|}$. This overlap ratio is 0 if $\tilde{E}_{S}$ and ${E}_{S}$ are exactly the same,  and 1 if they are completely different. 
Hence a larger ratio indicates a better attack performance. 
In our results, we report the average overlap ratio on a set of testing nodes and graphs. 

{\bf Parameter setting:} Table \ref{table:setting} in Appendix~\ref{supp:ExperimentSetting} summarizes the 
default values of the key parameters in the explainers and our attack, e.g., perturbation budget $\xi$, top-$k$ selection parameter $k$. 
When  studying the impact of each parameter, we fix the others to be  the default value. 

{\bf Compared attacks:}
There exist no attack methods on GNN explainers. Here, 
we propose two baselines for comparison.

\begin{table}[!t] \renewcommand{\arraystretch}{0.95}
	\centering
 \addtolength{\tabcolsep}{-3pt}
  \scriptsize
	\caption{Testing accuracy on the trained GCN model.}
 
	\begin{tabular}{ccccccc}
		\toprule 
		Model &House&Community&Cycle&OGBN-P&MUTAG&REDDIT\\ 
		\midrule 
		GCN&92.32\% &83.17\%&92.53\%&81.12\% &86.28\% &80.88\%\\
        \bottomrule 
	\end{tabular}
	\label{table:accuracy}
\vspace{-2mm}
\end{table}

\emph{Random attack:} This attack iteratively and randomly selects a pair of nodes.  
If there exists no edge between them, we add one; otherwise, if the existing edge is not from the original explanation, we delete it. 
Finally, we ensure the total added and deleted edges to be $\xi$ and pass the attack constraints, e.g., likelihood ratio test and maintain the accurate prediction. 

\emph{Kill-hot attack:} This attack assumes the adversary knows the mask generated by the explainer. It first decides the non-explanatory edges with the highest mask values and then deletes the top-$\xi$ edges, presuming these edges could heavily influence the mask.

\subsection{Experimental Results}

\begin{figure}[!t]
\begin{center}
    \captionsetup[subfloat]{labelsep=none,format=plain,labelformat=empty,farskip=0pt}
 \vspace{+2mm}
	\begin{minipage}[t]{\linewidth}
       \begin{center}
       
 \scriptsize
        \subfloat{
        \subfloat{
    	\includegraphics[width=0.133\linewidth,frame]{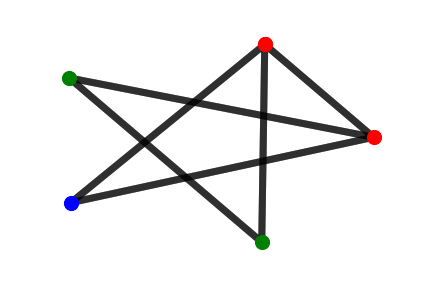}}
        \subfloat{
    	\includegraphics[width=0.133\linewidth,frame]{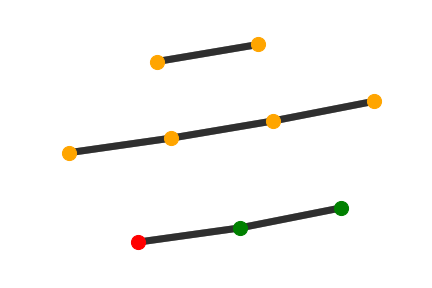}}}
        \quad
        \subfloat{
        \subfloat{
    	\includegraphics[width=0.133\linewidth,frame]{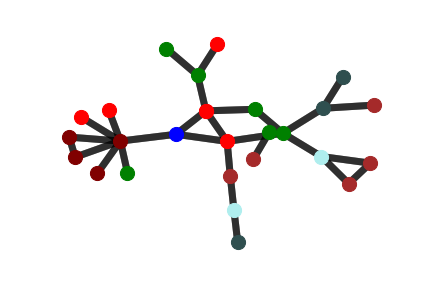}}
        \subfloat{
    	\includegraphics[width=0.133\linewidth,frame]{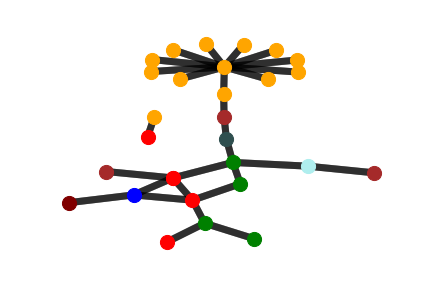}}}
        \quad
        \subfloat{
        \subfloat{
    	\includegraphics[width=0.133\linewidth,frame]{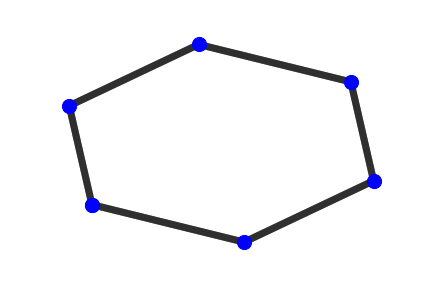}}
        \subfloat{
    	\includegraphics[width=0.133\linewidth,frame]{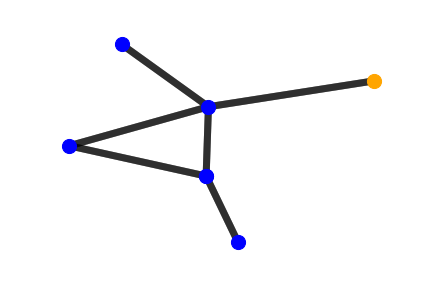}}}
        \end{center}
	\end{minipage}	
 
    \begin{minipage}[t]{\linewidth}
       \begin{center}
       
 \scriptsize
        \subfloat{
        \subfloat{
    	\includegraphics[width=0.133\linewidth,frame]{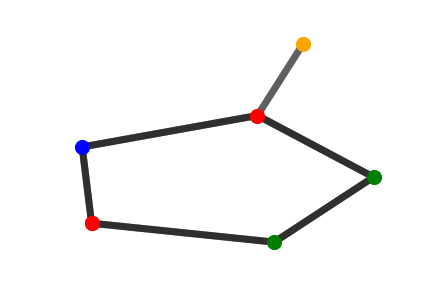}}
        \subfloat{
    	\includegraphics[width=0.133\linewidth,frame]{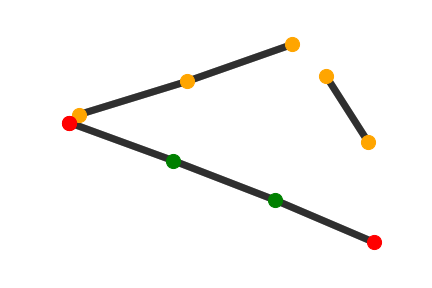}}}
        \quad
        \subfloat{
        \subfloat{
    	\includegraphics[width=0.133\linewidth,frame]{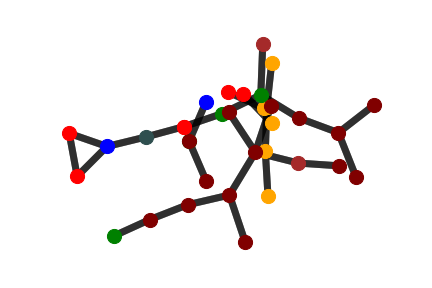}}
        \subfloat{
    	\includegraphics[width=0.133\linewidth,frame]{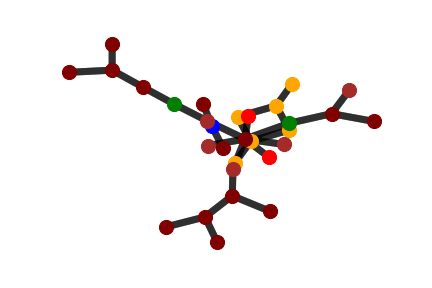}}}
        \quad
        \subfloat{
        \subfloat{
    	\includegraphics[width=0.133\linewidth,frame]{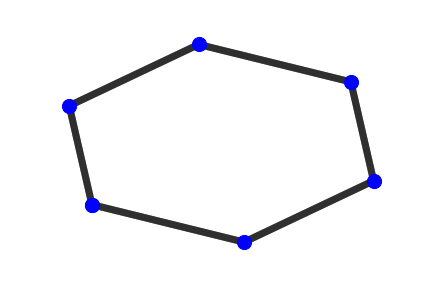}}
        \subfloat{
    	\includegraphics[width=0.133\linewidth,frame]{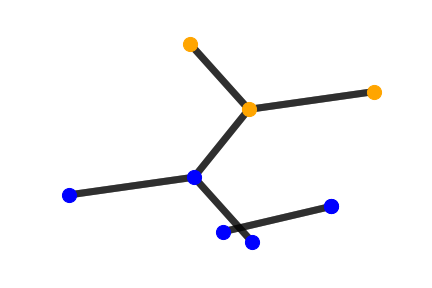}}}
        \end{center}
	\end{minipage}
    \begin{minipage}[t]{\linewidth}
       \begin{center}
       
 \scriptsize
        \subfloat{
        \subfloat{
    	\includegraphics[width=0.133\linewidth,frame]{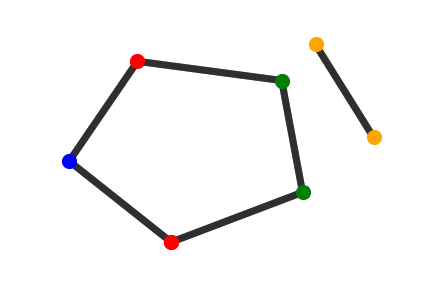}}
        \subfloat{
    	\includegraphics[width=0.133\linewidth,frame]{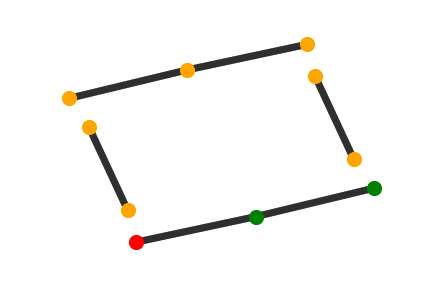}}}
        \quad
        \subfloat{
        \subfloat{
    	\includegraphics[width=0.133\linewidth,frame]{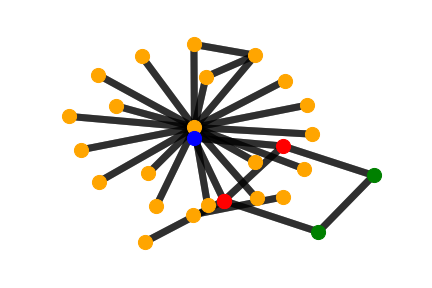}}
        \subfloat{
    	\includegraphics[width=0.133\linewidth,frame]{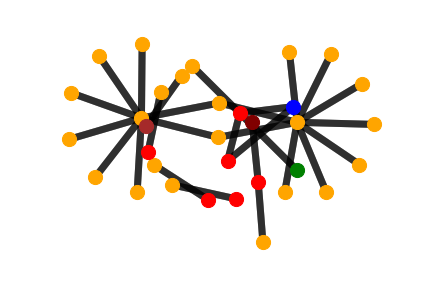}}}
        \quad
        \subfloat{
        \subfloat{
    	\includegraphics[width=0.133\linewidth,frame]{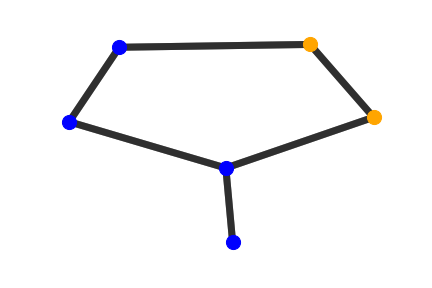}}
        \subfloat{
    	\includegraphics[width=0.133\linewidth,frame]{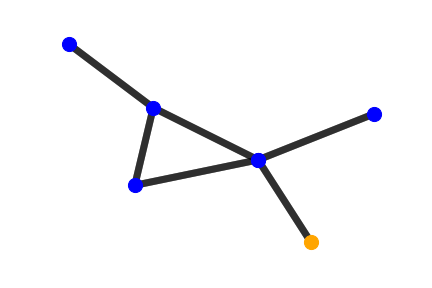}}}
        \end{center}
	\end{minipage}
    \begin{minipage}[t]{\linewidth}
    
 \scriptsize
       \begin{center}
        \subfloat[
 \tiny House]{
        \subfloat[\scriptsize $E_{S}$]{
    	\includegraphics[width=0.133\linewidth,frame]{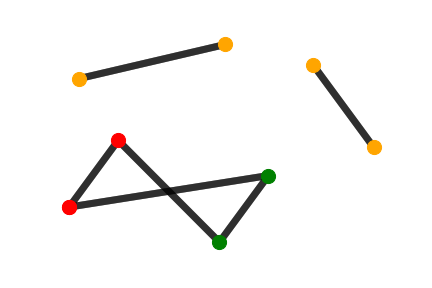}}
        \subfloat[\scriptsize$\tilde{E}_S$]{
    	\includegraphics[width=0.133\linewidth,frame]{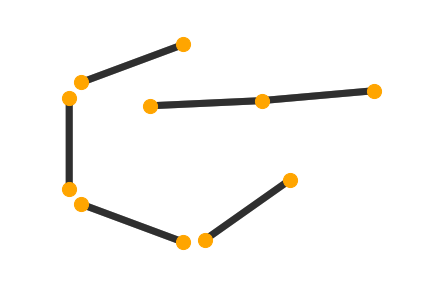}}}
        \quad
        \subfloat[
 \tiny Community]{
        \subfloat[\scriptsize $E_{S}$]{
    	\includegraphics[width=0.133\linewidth,frame]{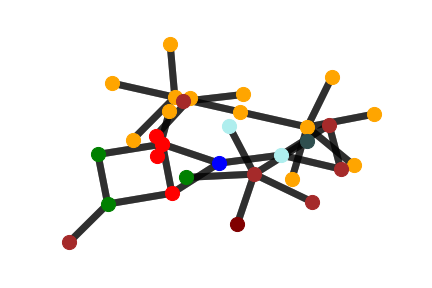}}
        \subfloat[\scriptsize$\tilde{E}_{S}$]{
    	\includegraphics[width=0.133\linewidth,frame]{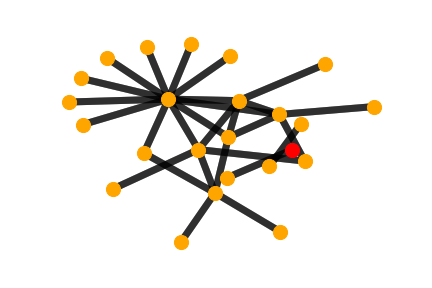}}}
        \quad
        \subfloat[
 \tiny Cycle]{
        \subfloat[\scriptsize $E_{S}$]{
    	\includegraphics[width=0.133\linewidth,frame]{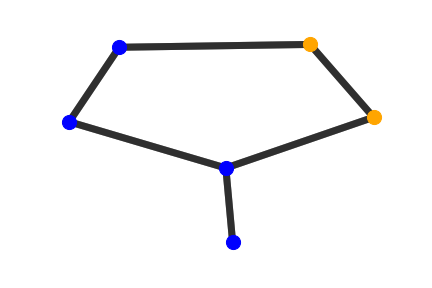}}
        \subfloat[\scriptsize$\tilde{E}_S$]{
    	\includegraphics[width=0.133\linewidth,frame]{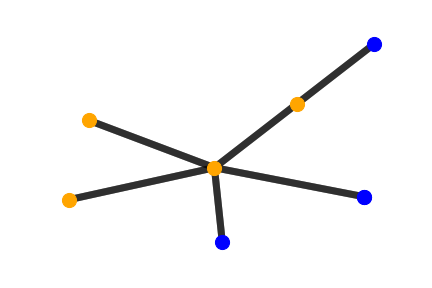}}}
        \end{center}
	\end{minipage} 

	\begin{minipage}[t]{\linewidth}
 \scriptsize
       \begin{center}
       
        \subfloat{
        \subfloat{
    	\includegraphics[width=0.133\linewidth,frame]{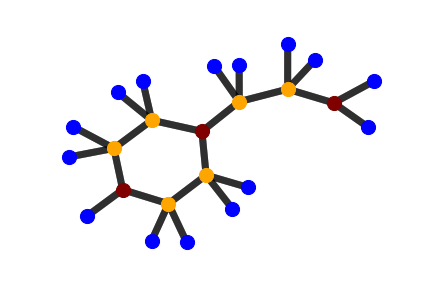}}
        \subfloat{
    	\includegraphics[width=0.133\linewidth,frame]{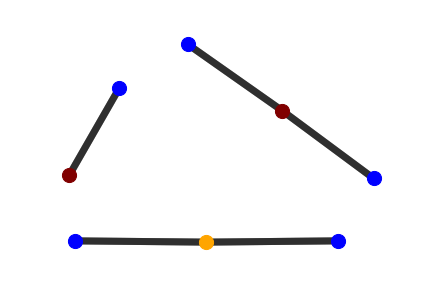}}
    
        \subfloat{
    	\includegraphics[width=0.133\linewidth,frame]{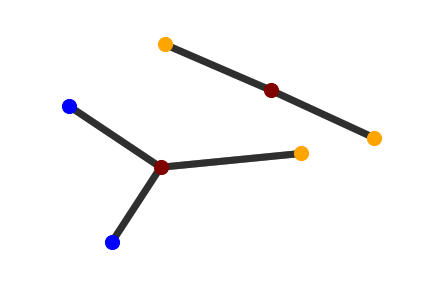}}}
        \quad
        \quad
        \subfloat{
        \subfloat{
    	\includegraphics[width=0.133\linewidth,frame]{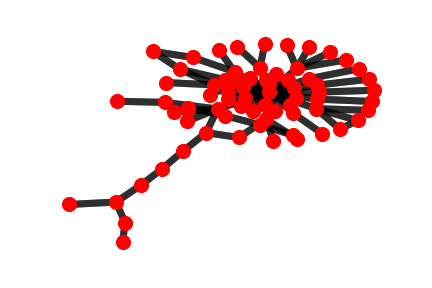}}
        \subfloat{
    	\includegraphics[width=0.133\linewidth,frame]{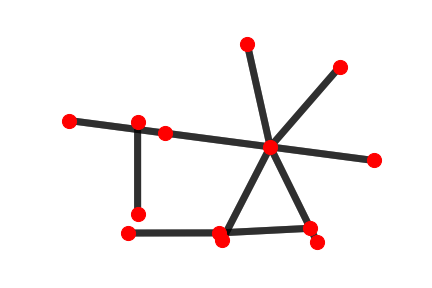}}
        \subfloat{
    	\includegraphics[width=0.133\linewidth,frame]{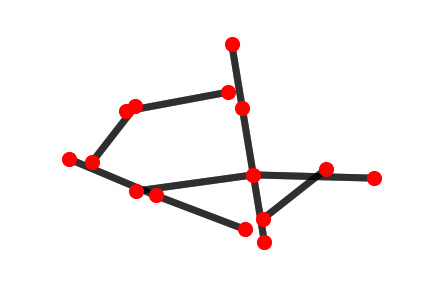}}}
        \end{center}
	\end{minipage}	
	\begin{minipage}[t]{\linewidth}
 \scriptsize
       \begin{center}
       
        \subfloat{
        \subfloat{
    	\includegraphics[width=0.133\linewidth,frame]{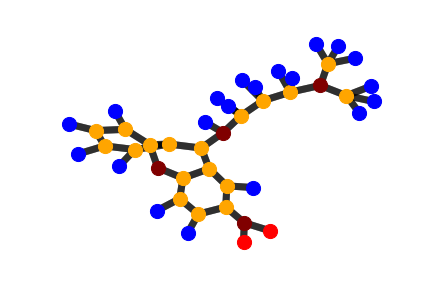}}
        \subfloat{
    	\includegraphics[width=0.133\linewidth,frame]{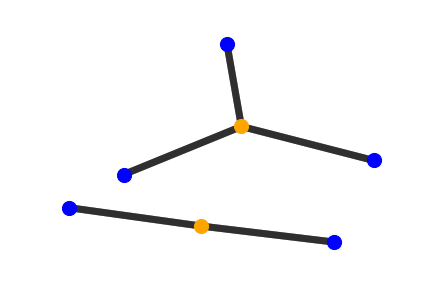}}
    
        \subfloat{
    	\includegraphics[width=0.133\linewidth,frame]{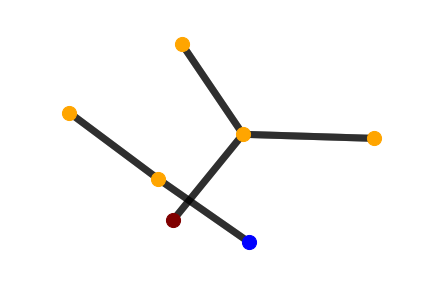}}}
        \quad
        \quad
        \subfloat{
        \subfloat{
    	\includegraphics[width=0.133\linewidth,frame]{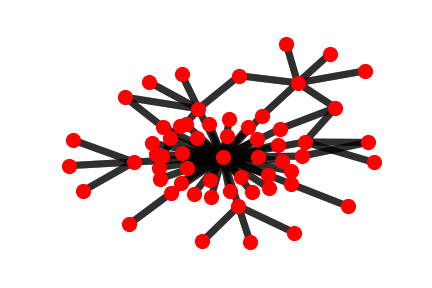}}
        \subfloat{
    	\includegraphics[width=0.133\linewidth,frame]{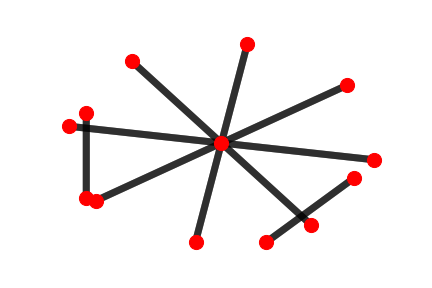}}
        \subfloat{
    	\includegraphics[width=0.133\linewidth,frame]{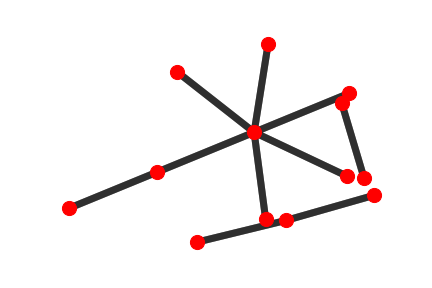}}}
        \end{center}
	\end{minipage}	
 	\begin{minipage}[t]{\linewidth}
  
 \scriptsize
       \begin{center}
       
        \subfloat{
        \subfloat{
    	\includegraphics[width=0.133\linewidth,frame]{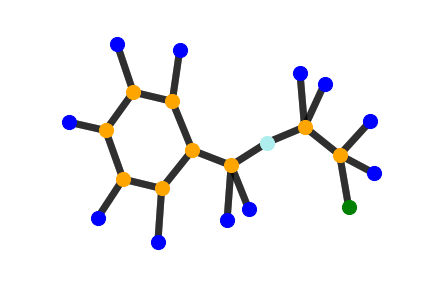}}
        \subfloat{
    	\includegraphics[width=0.133\linewidth,frame]{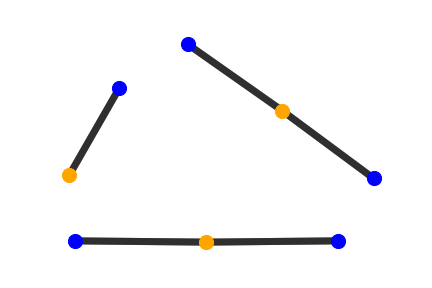}}
    
        \subfloat{
    	\includegraphics[width=0.133\linewidth,frame]{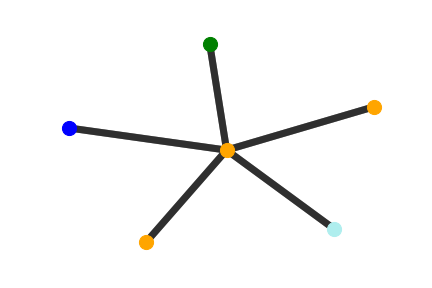}}}
        \quad
        \quad
        \subfloat{
        \subfloat{
    	\includegraphics[width=0.133\linewidth,frame]{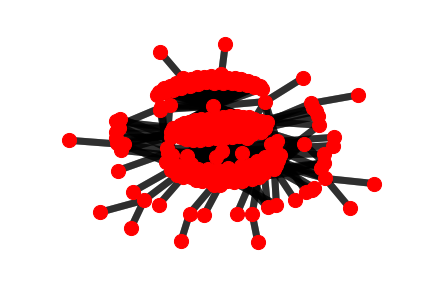}}
        \subfloat{
    	\includegraphics[width=0.133\linewidth,frame]{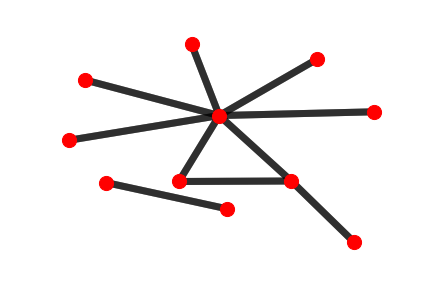}}
        \subfloat{
    	\includegraphics[width=0.133\linewidth,frame]{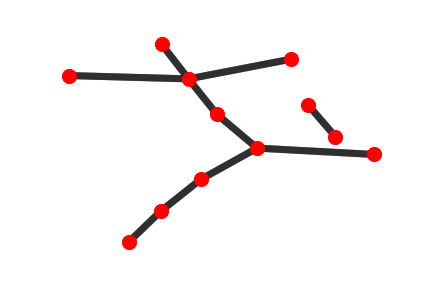}}}
        \end{center}
	\end{minipage}	
    \begin{minipage}[t]{\linewidth}
    
 \scriptsize
       \begin{center}
        \subfloat[
 \tiny MUTAG]{
        \subfloat[\scriptsize $E$]{
    	\includegraphics[width=0.133\linewidth,frame]{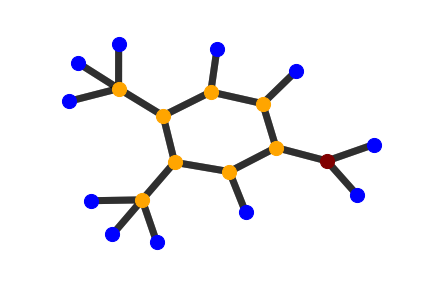}}
        \subfloat[\scriptsize ${E}_S$]{
    	\includegraphics[width=0.133\linewidth,frame]{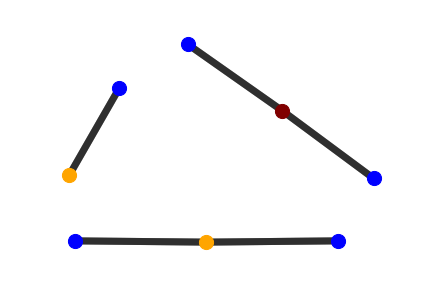}}
        \subfloat[\scriptsize$\tilde{E}_S$]{
    	\includegraphics[width=0.133\linewidth,frame]{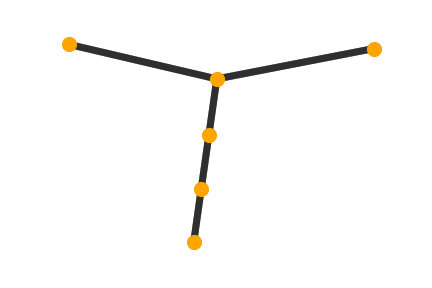}}}
        \quad
        \quad
        \subfloat[
 \tiny REDDIT]{
        \subfloat[\scriptsize $E$]{
    	\includegraphics[width=0.133\linewidth,frame]{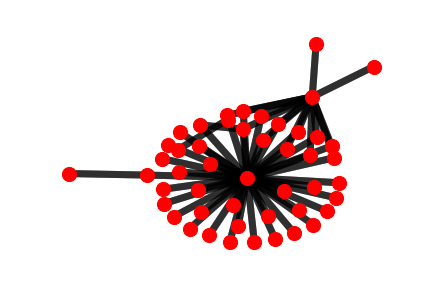}}
        \subfloat[\scriptsize $E_{S}$]{
    	\includegraphics[width=0.133\linewidth,frame]{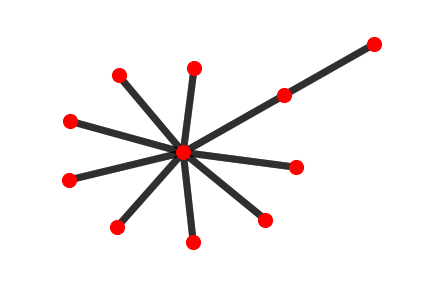}}
        \subfloat[\scriptsize $\tilde{E}_S$]{
    	\includegraphics[width=0.133\linewidth,frame]{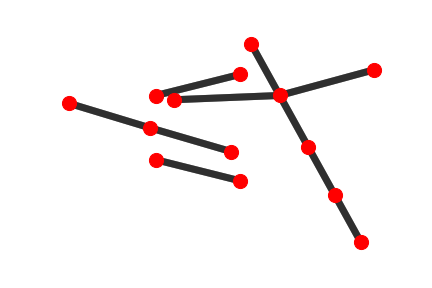}}}
        \end{center}
	\end{minipage} 
  \vspace{-2mm}
	\caption{Visualization of the explanation results before and after our deduction-based attack against the PGExplainer. We do not show OGBN-P as it is too big/dense to be visualized. 
 }
	\label{fig:Visualize}
\vspace{-6mm}
\end{center}
\end{figure}

\begin{table}[!t]\renewcommand{\arraystretch}{0.9}
\scriptsize	
 \centering
\addtolength{\tabcolsep}{-5.5pt}
\caption{Overall attack performance.}
	\begin{tabular}{cccccccc}
		\toprule  
		Explainer& Method &House&Community&Cycle&OGBN-P&MUTAG&REDDIT\\ 
		\midrule  
		&Random&5.04\%&16.75\%&22.99\%&21.09\%&34.55\%&33.26\%\\
		GNNExp. & Kill-hot&11.11\%&18.25\%&14.77\%&17.62\%&42.40\%&34.40\%\\
		 & Loss-based&{19.05\%}&{28.35\%}&{57.92\%}&40.62\%   &{63.40\%}&{47.96\%}\\
		 & Dedu.-based&\textbf{20.02}\%&\textbf{29.79\%}&\textbf{64.42\%}&\textbf{42.15\%}&\textbf{64.80\%}&\textbf{48.07\%}\\
		\midrule 
		&Random&16.55\%&10.50\%&5.30\%&7.72\%&44.66\%&29.72\%\\
		PGExp. & Kill-hot&14.50\%&13.76\%&5.46\%&7.29\%&55.60\%&29.55\%\\
		 & Loss-based&35.83\%&24.46\%&37.59\%&35.65\%&\textbf{64.80\%}&{42.63\%}\\
		& Dedu.-based&{\textbf{41.50\%}}&{\textbf{24.59\%}}&{\textbf{47.93\%}}&\textbf{36.12\%
}&\textbf{64.80\%}&\textbf{43.05\%}\\
		\midrule 
		&Random&34.04\%&13.72\%&9.30\%&11.94
\%&26.08\%&26.78\%\\
		GSAT & Kill-hot&36.92\%&13.36\%&8.33\%&8.12
\%&26.50\%&28.05\%\\
		 & Loss-based&\textbf{51.25\%}&20.20\%&32.33\%&30.82\%&{68.50\%}&{50.25\%}\\
		 & Dedu.-based&\textbf{51.25\%}&\textbf{45.54\%}&\textbf{48.45\%}& \textbf{31.88
\%}    &\textbf{69.90\%}&\textbf{57.40\%}\\
        \bottomrule  
	\end{tabular}
	\label{table:result}
 \vspace{-2mm}
\end{table}

{\bf Overall results:} Table~\ref{table:result} shows the attack performance of the compared attacks on the three GNN explainers and six datasets. 
We have several key observations: 1) Random attack performs the worst, indicating randomly selecting edges for perturbation is not effective enough. 2) Kill-hot attack performs (slightly) better than random attack in most cases, but much worse than our two attacks. This implies the ``most important" non-explanatory edges contain some information to influence the final mask, but is not sufficient; 3) Deduction-based attack 
performs the best and loss-based attack performs the second best. 
This shows loss change is a good indicator to identify certain important edges for perturbation, while  
simulating the learning procedure of (black-box) GNN explainers can be more beneficial to identify the important edges; 
4) We do not see a strong connection between the attack performance on explainers and GNN's testing accuracy, by comparing Table~\ref{table:result} and Table~\ref{table:accuracy}.  

Note that when running our attacks, we filter the intermediate perturbed graphs whose predictions are not maintained. Here we also report the fraction of GNN mis-predictions during the attack in Table~\ref{table:failed}. We can see the fraction is low, indicating the perturbed graph does not change the prediction often. The main reason is that we focus more on attacking the GNN explanations, instead of attacking the GNN predictions like, e.g., \citet{fan2022jointlyAttackExplanation}.

{\bf Visualizing explanation results:} 
To better understand the vulnerability caused by our attacks, we visualize the explanation results
without and with our deduction-based attack on 
 PGExplainer in Figure~\ref{fig:Visualize}
 We can categorize these results into two types: 1) Both explanatory edges obtained by the explainer without attack and those obtained with our attack are in a connected subgraph, but they look significantly different;  
2) Explanatory edges obtained by the explainer without attack are in a connected subgraph, but they are disconnected after our attack.

\begin{table}[!t] \renewcommand{\arraystretch}{0.95}
\centering
\scriptsize	
 \addtolength{\tabcolsep}{-5pt}
	\caption{Fraction of failed attacks during training.}
 	\begin{tabular}{cccccccc}
		\toprule 
		Explainer& Method &House&Community&Cycle&OGBN-P&MUTAG&REDDIT\\ 
		\midrule  
		GNNExp. & Loss-based&{4.2\%}&{14.1\%}&{11.4\%}&4.6\%   &{1.7\%}&{0.0\%}\\
		& Dedu.-based&7.5\%&{16.5\%}&{5.0\%}&{7.7\%}&{0.0\%}&{0.0\%}\\
		\midrule 
		PGExp. & Loss-based&2.9\%&12.6\%&5.0\%&0.0\%&{4.2\%}&{0.0\%}\\
		 & Dedu.-based&{{2.1\%}}&{{9.2\%}}&{{10.4\%}}&{0.0\%
}&{0.0\%}&{0.0\%}\\
		\midrule 
		GSAT & Loss-based& {3.8\%}&12.6\%&6.7\%&0.0\%&{0.0\%}&{0.0\%}\\
		& Dedu.-based&{2.1\%}&{9.6\%}&{8.3\%}&{0.0\%}&{1.7\%}&{0.0\%}\\	
        \bottomrule 
	\end{tabular}
	\label{table:failed}
\end{table}

\begin{figure}[!t]
    \captionsetup[subfloat]{labelsep=none,format=plain,labelformat=empty,farskip=0pt}
    \centering
	\begin{minipage}[t]{\linewidth}
 \begin{tiny}
        \begin{center}
        \subfloat[
 \tiny OGBN-P]{
		\includegraphics[width=0.32\linewidth]{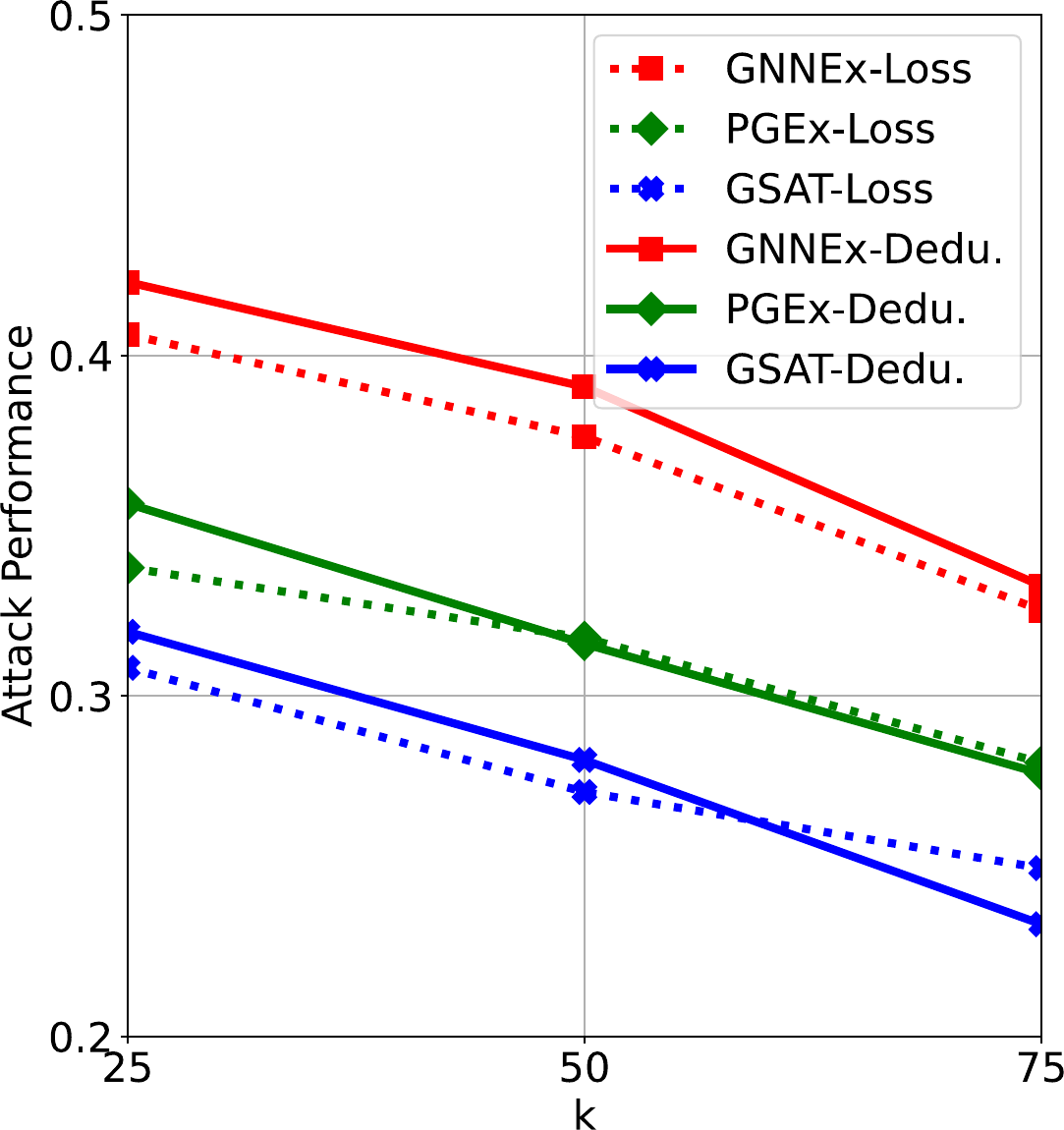}	}
        \subfloat[
 \tiny MUTAG]{
        \includegraphics[width=0.32\linewidth]{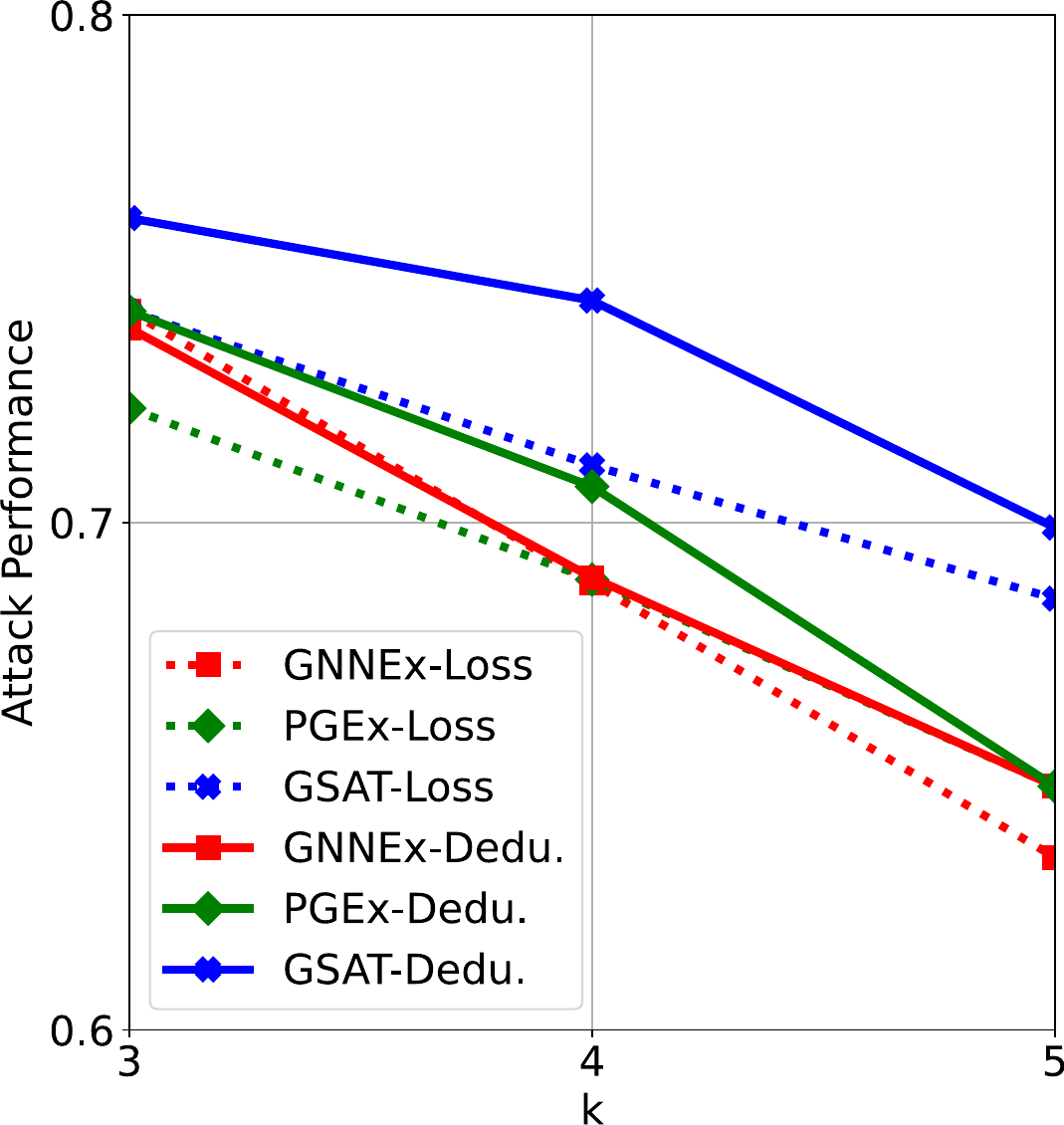}} \subfloat[
 \tiny REDDIT]{
        \includegraphics[width=0.32\linewidth]{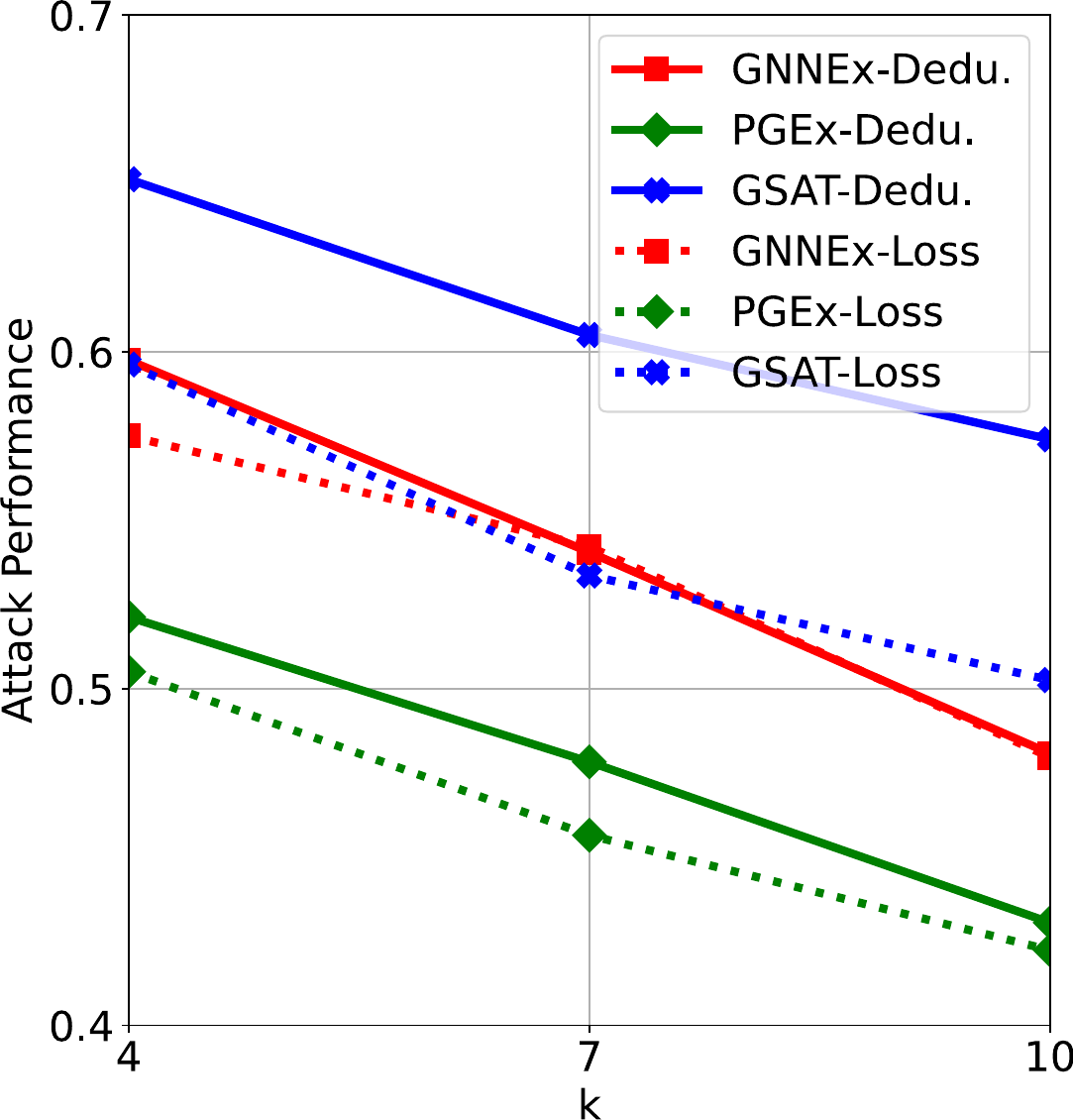}}
        \end{center}
        \begin{center}\footnotesize
        \end{center}
        \end{tiny}
	\end{minipage}
  \vspace{-6mm}
    \caption{{Impact of $k$ on our attack in three real-world datasets. }}
    \label{fig:ImpactK}
\vspace{-2mm}
\end{figure}

\begin{figure}[!t]
    \captionsetup[subfloat]{labelsep=none,format=plain,labelformat=empty,farskip=0pt}
    \centering
 \begin{minipage}[t]{\linewidth}
        \begin{center}
        \subfloat[\tiny(a) Impact of $N$]{
        \includegraphics[width=0.32\linewidth]{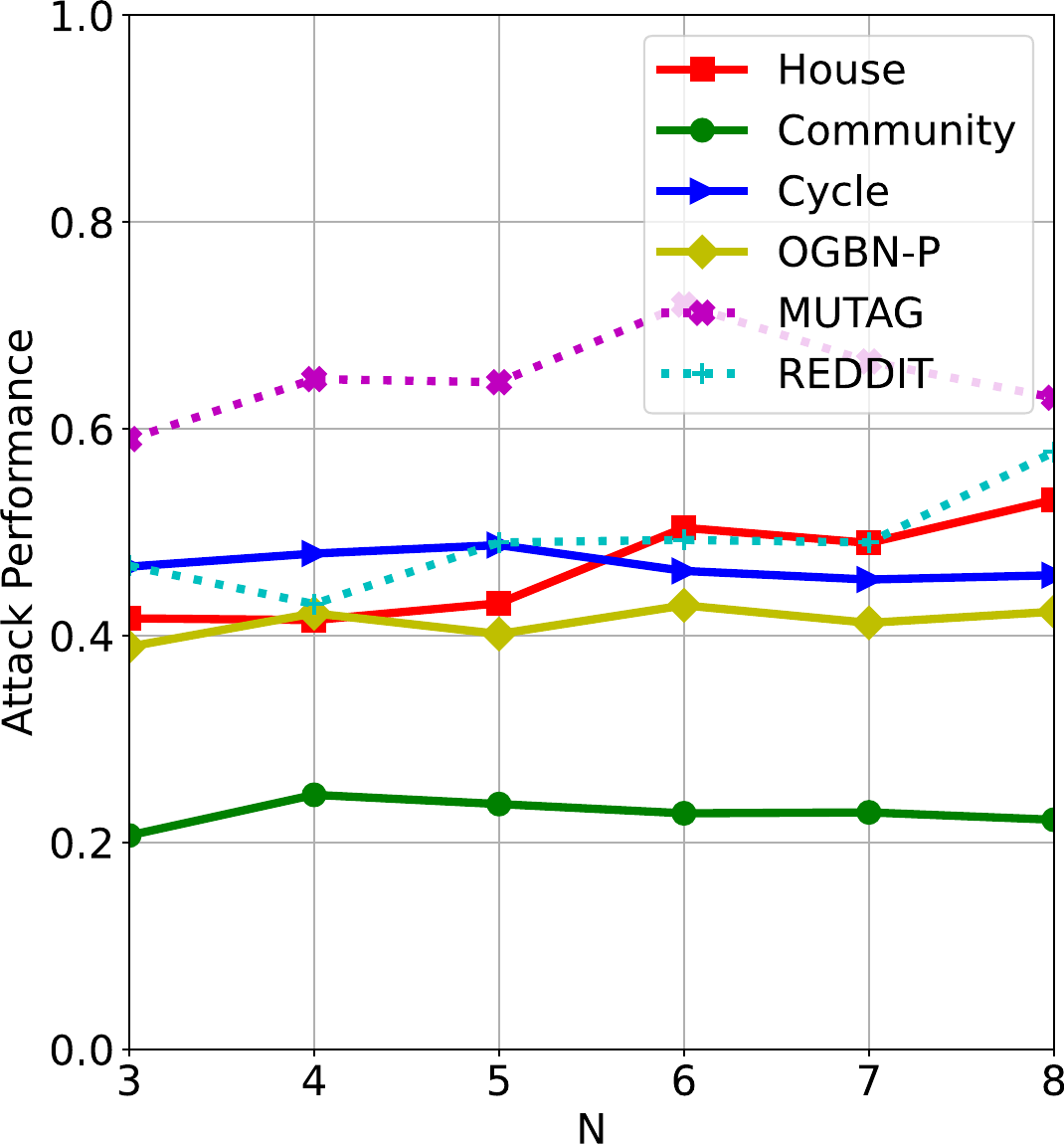}} 
        \subfloat[\tiny(b) Impact of $\gamma$ (left) and $\beta$ (right)]{
		\includegraphics[width=0.32\linewidth]{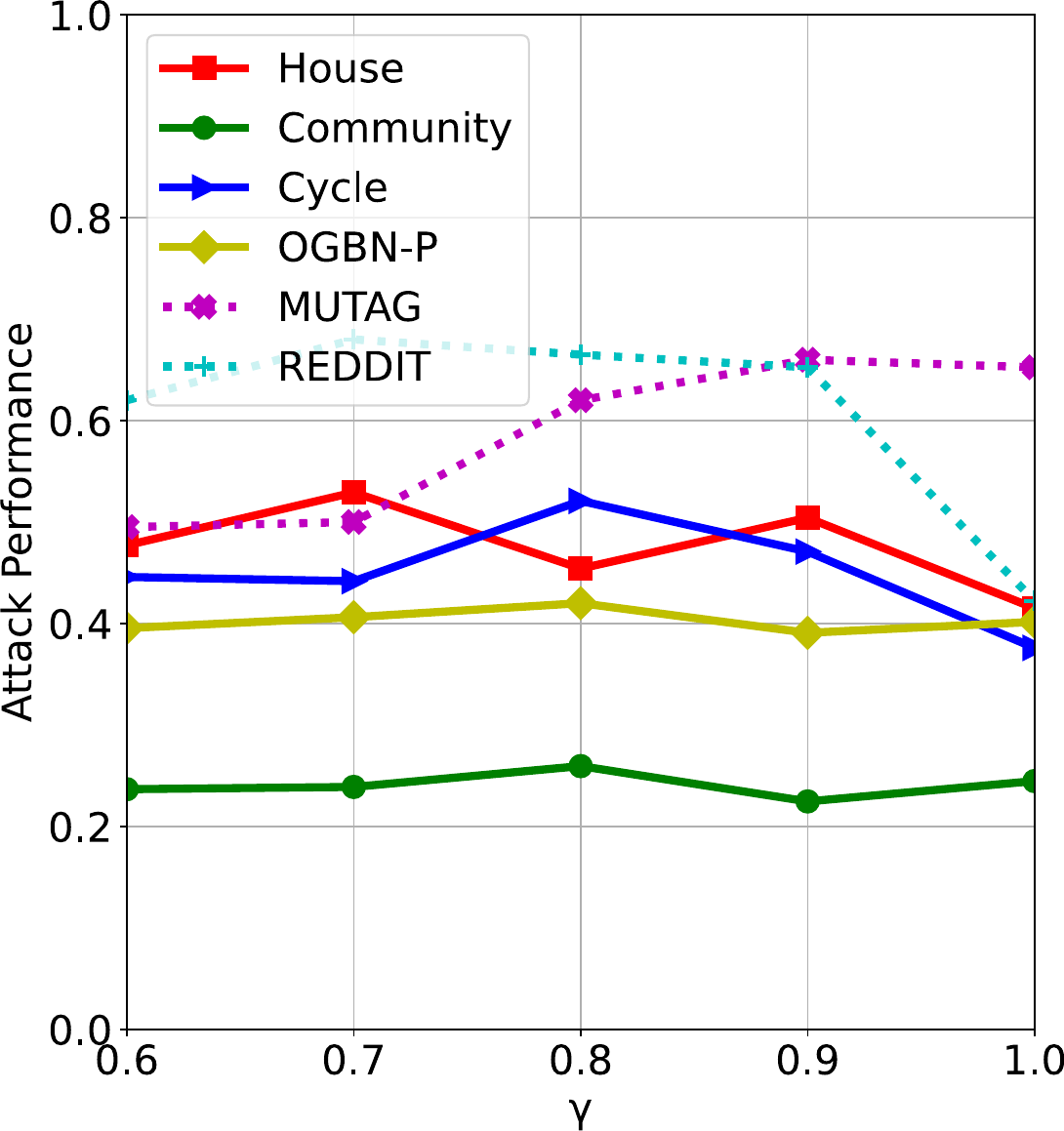}
		\includegraphics[width=0.32\linewidth]{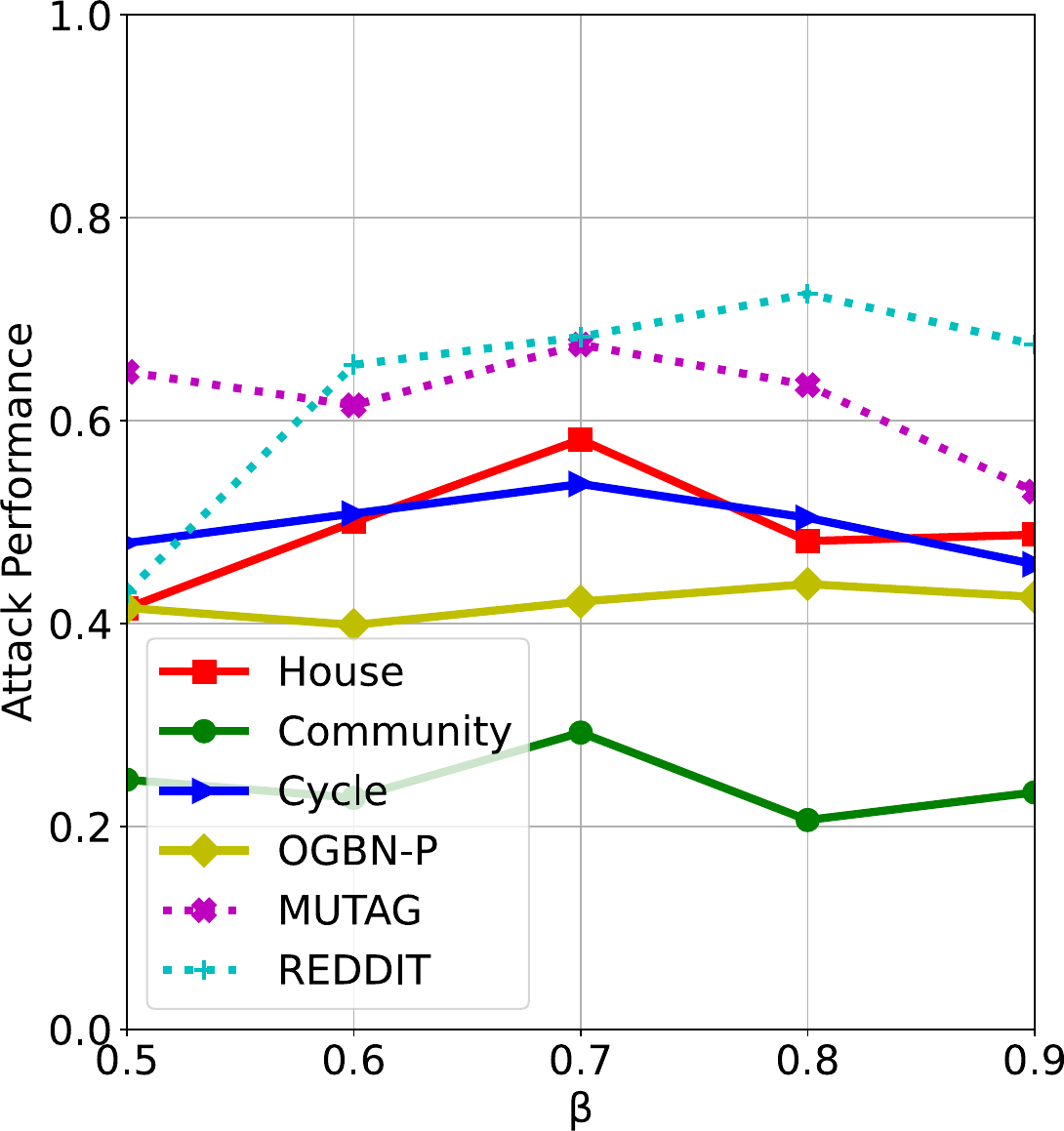}}	
        \end{center}
        \begin{center}\footnotesize
        \end{center}
	\end{minipage}
 \vspace{-6mm}
    \caption{(a) Impact of $N$ on our deduction-based attack; (b) {Impact of $\gamma$ and $\beta$ on our attack performance.}}
    \label{fig:ImpactN}
    \vspace{-4mm}
\end{figure}

\begin{figure*}[!t]
    \captionsetup[subfloat]{labelsep=none,format=plain,labelformat=empty,farskip=0pt}
    \centering
	\begin{minipage}[t]{\linewidth}
 \begin{tiny}
        \begin{center}
        \subfloat[\tiny GNNExplainer]{
        \subfloat[
 \tiny Loss-based ]{
		\includegraphics[width=0.16\linewidth]{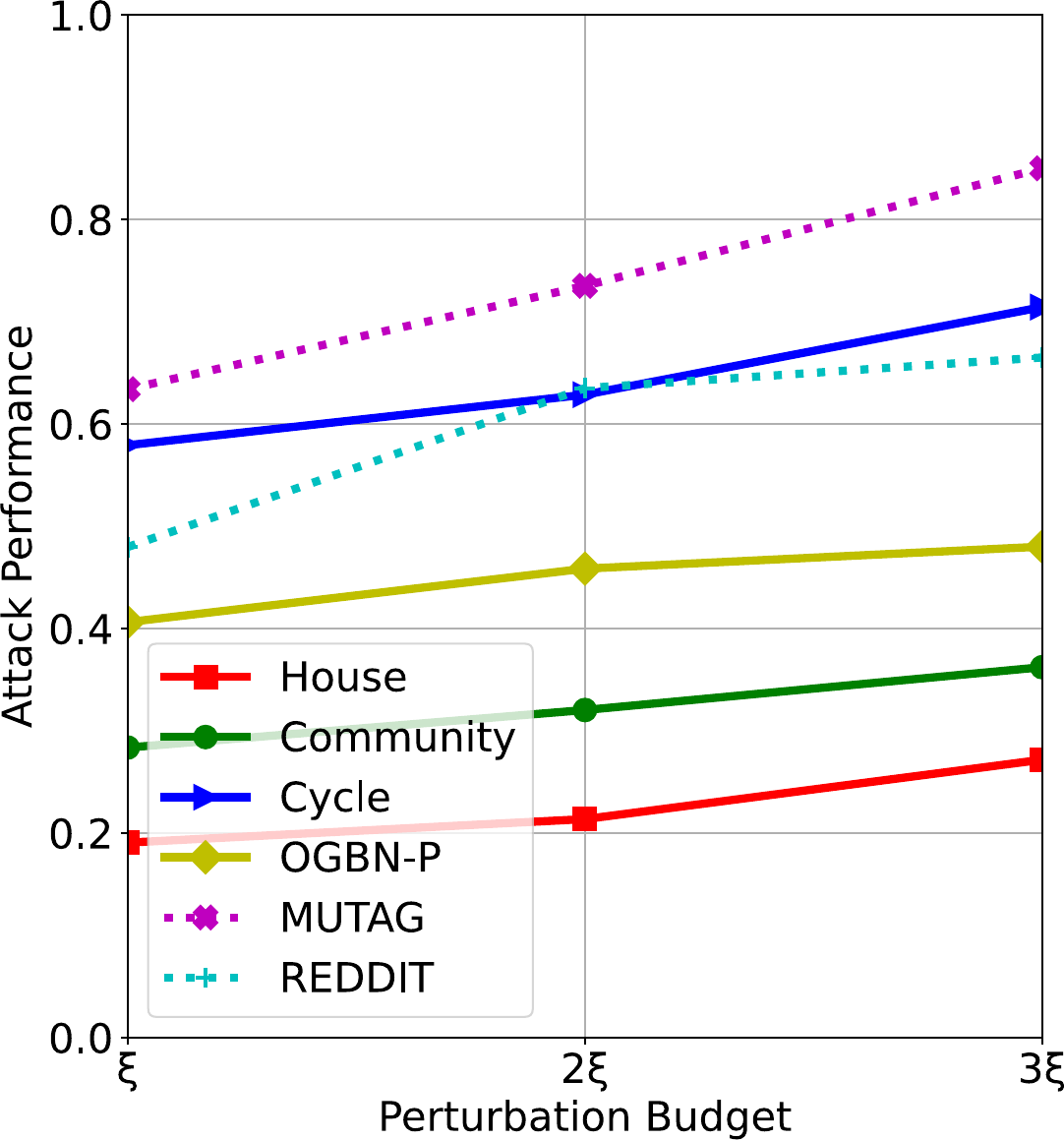}	}
        \subfloat[
 \tiny Dedu.-based]{
        \includegraphics[width=0.16\linewidth]{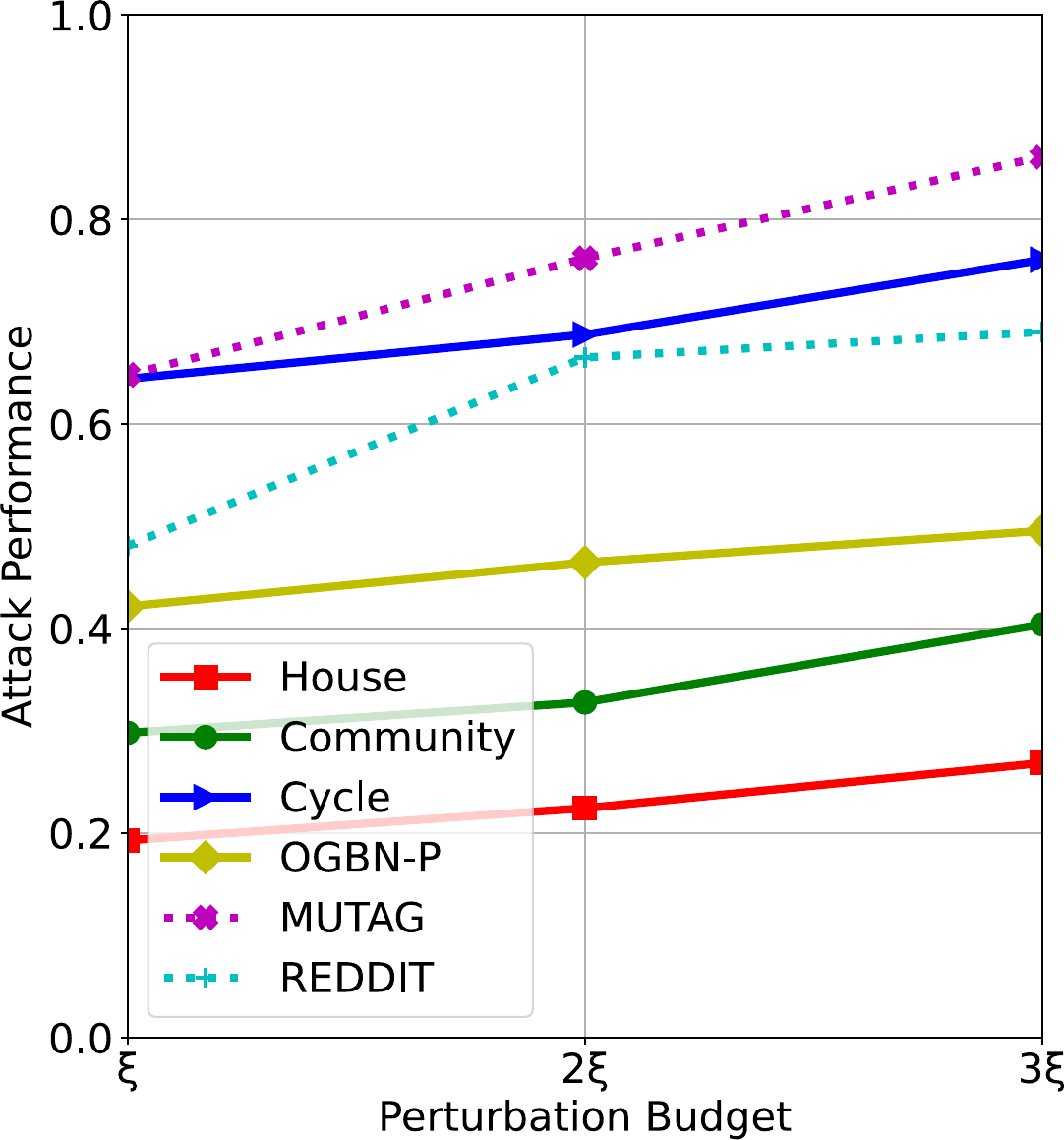}} }
        \subfloat[\tiny PGExplainer]{
        \subfloat[
 \tiny Loss-based ]{
		\includegraphics[width=0.16\linewidth]{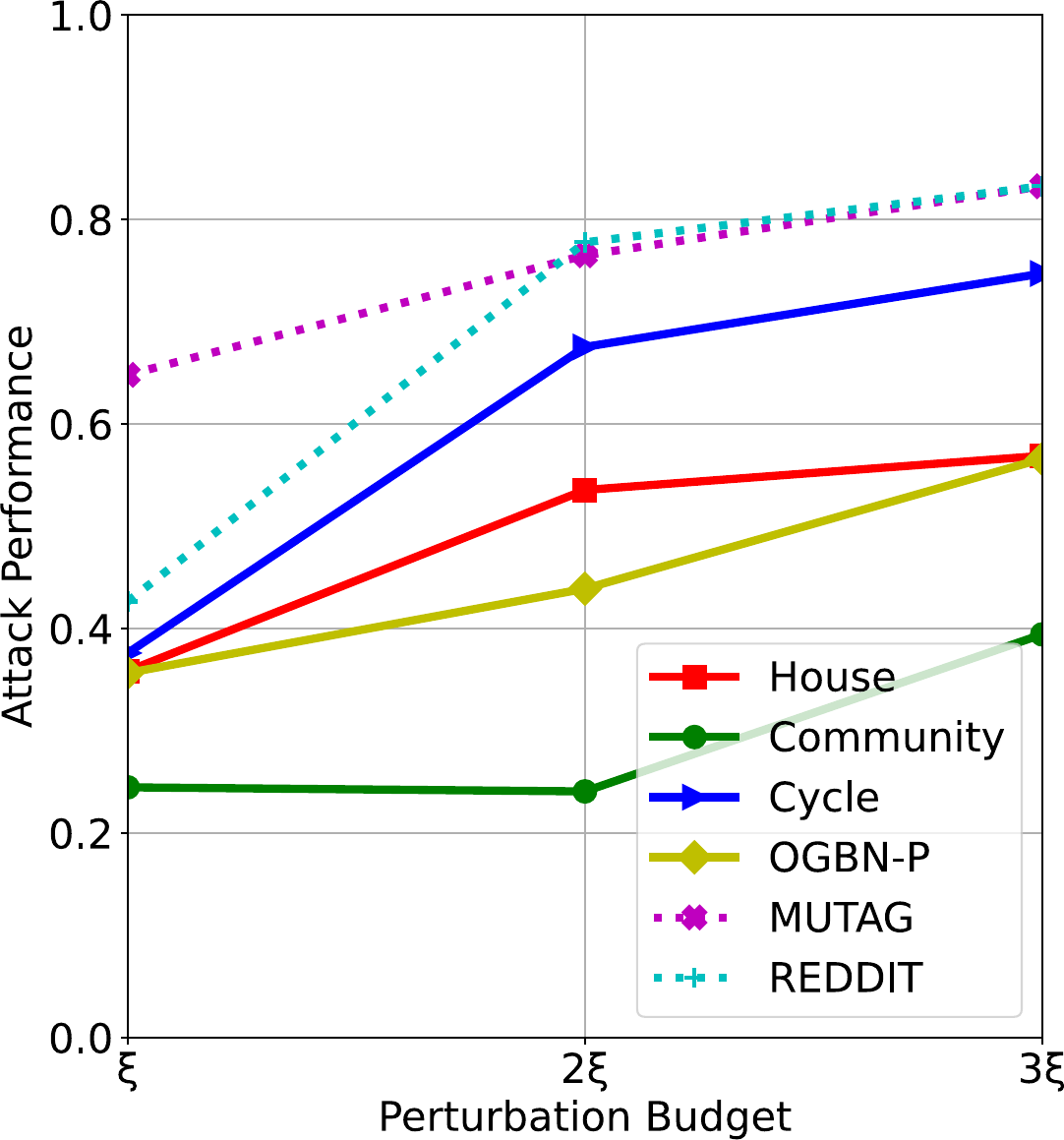}	}
        \subfloat[
 \tiny Dedu.-based]{
        \includegraphics[width=0.16\linewidth]{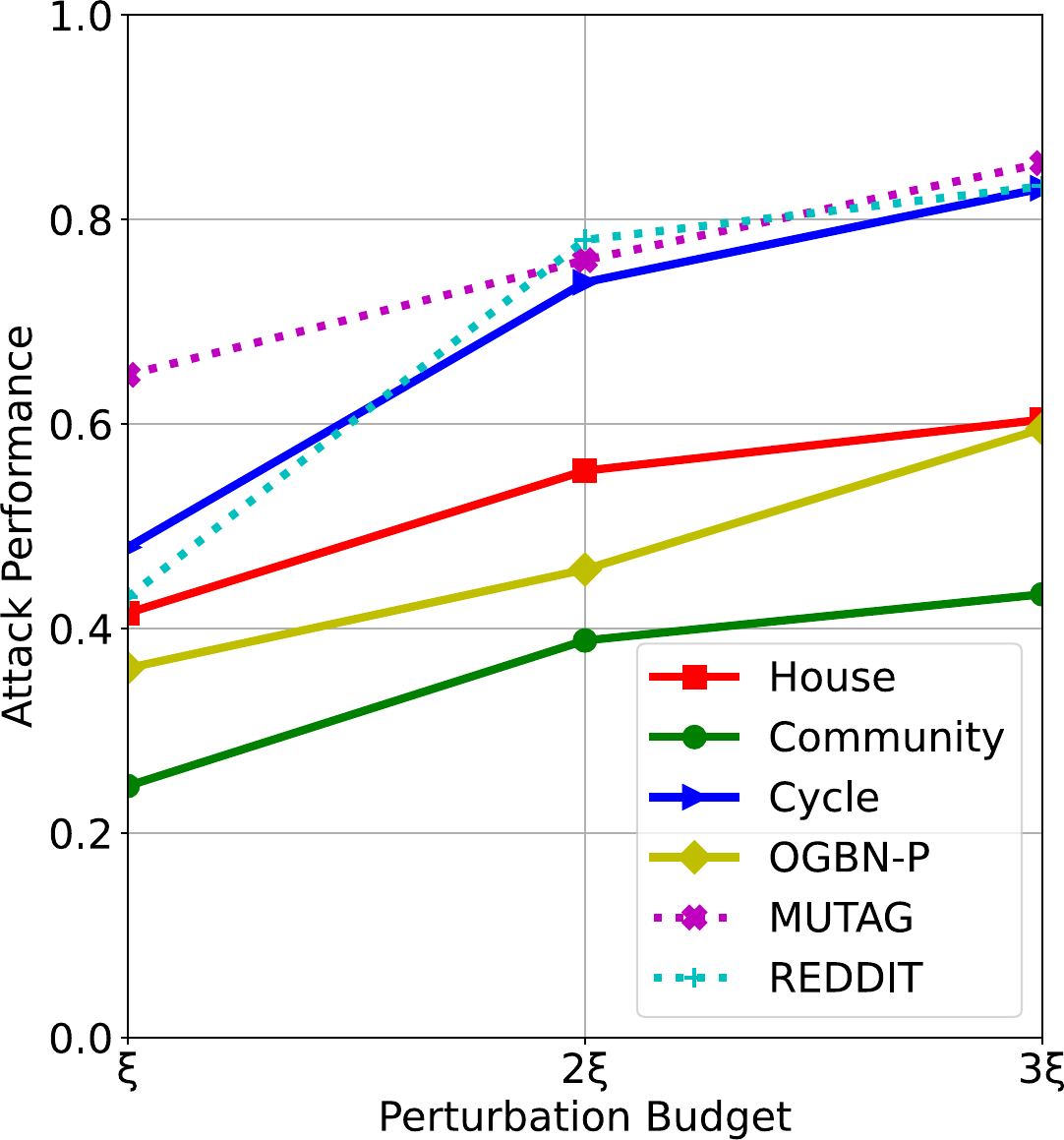}} }
        \subfloat[\tiny GSAT]{
        \subfloat[
 \tiny Loss-based ]{
		\includegraphics[width=0.16\linewidth]{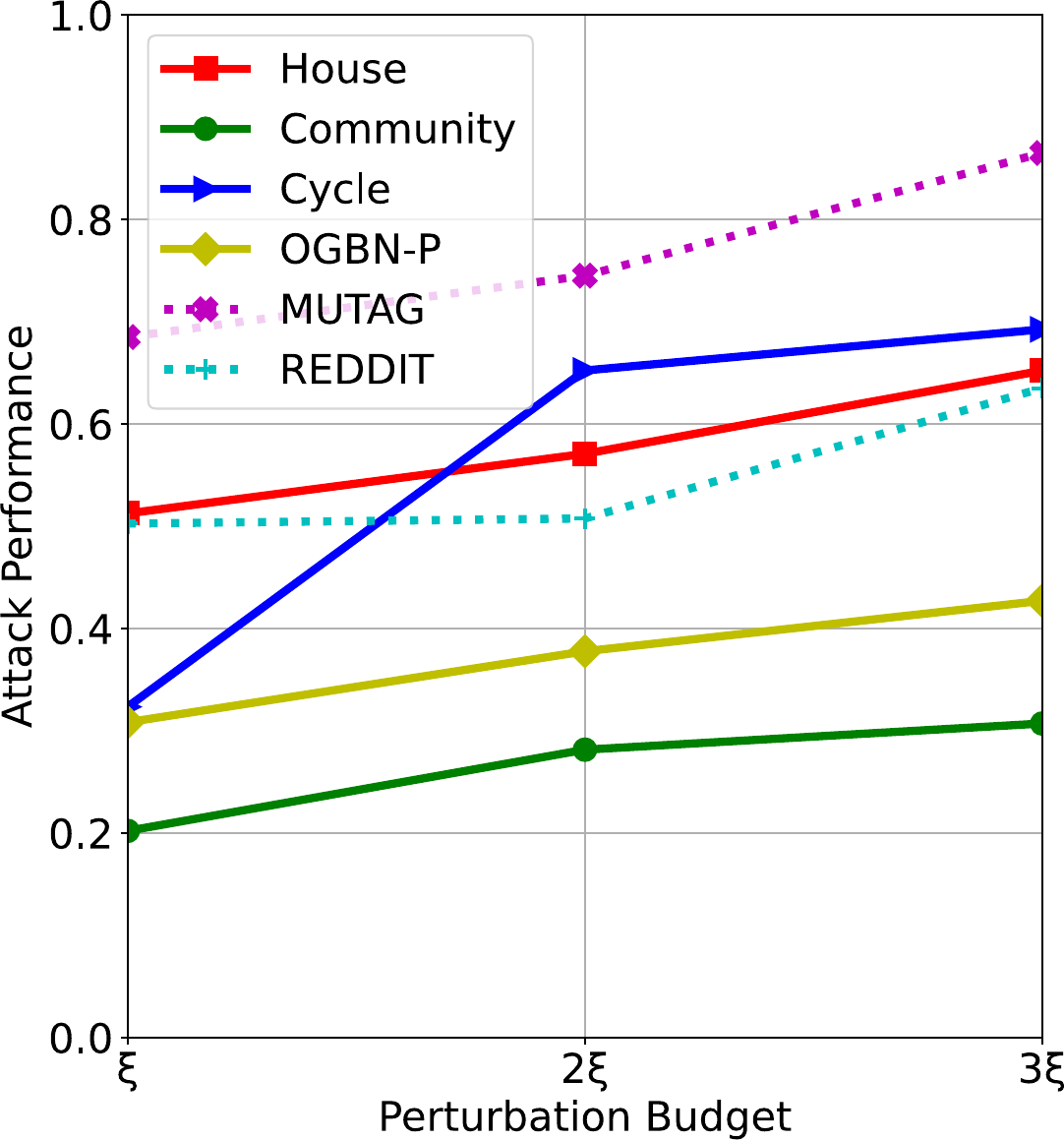}	}
        \subfloat[
 \tiny Dedu.-based]{
        \includegraphics[width=0.16\linewidth]{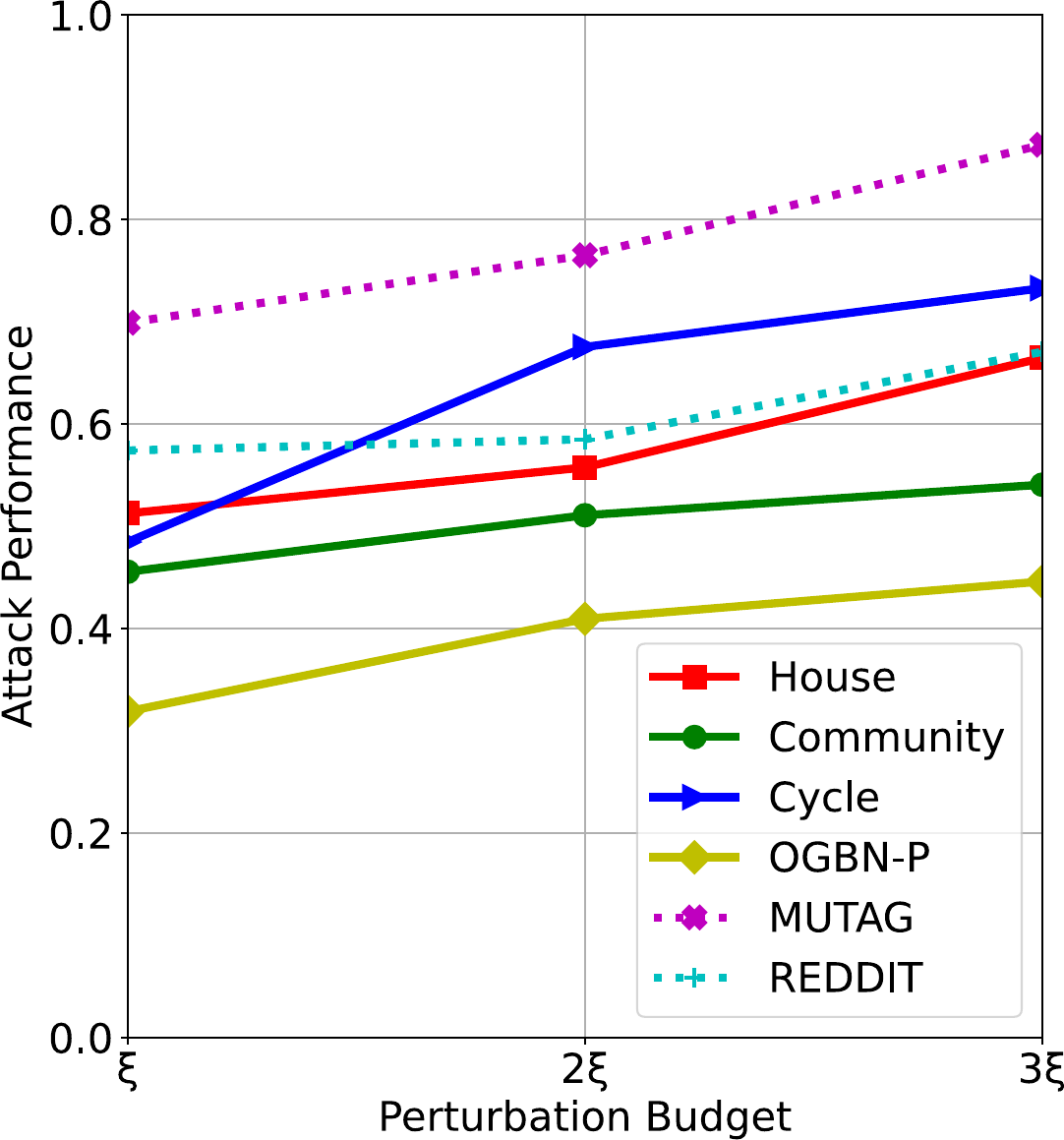}} }
        
        \end{center}
        \begin{center}\footnotesize
        \end{center}
        \end{tiny}
	\end{minipage}
  \vspace{-6mm}
    \caption{{Impact of $\xi$ on our attack performance.}}
    \label{fig:impactxi}
\vspace{-2mm}
\end{figure*}

{\bf Ablation study:} In this experiment, we show the impact of key parameters in our attack on the attack performance. 

\emph{Impact of $k$:} As synthetic graphs have a fixed size of the groundtruth explanation (e.g., $k=6$ in BA House), we only study the impact of $k$ (the size of the explanation results) on the real-world MUTAG and REDDIT graphs. The results are shown in Figure~\ref{fig:ImpactK}. We can see our attacks perform better on a smaller $k$. This is understandable, as it would be naturally more challenging to affect larger number of edges, with a limited perturbation budget $\xi=2$.  

\emph{Impact of $N$:} Our deduction-based attack uses a finite number of $N$ samples to approximate the integral. 
We analyze the effect of $N$ and show the results in Figure \ref{fig:ImpactN}(a). Overall, the attack results are stable with 
different $N$'s, e.g., the performance difference is within $5\%$. A larger $N$ shows slightly better performance. One reason could be that larger $N$ yields more accurate approximation of the integral.  

\emph{Impact of $\gamma$ and $\beta$:} Both our attacks force the mask values on the explanatory edges by the explainer without attack to be a constant ($\gamma$ or dependent on $\beta$) during attack optimization. Here we study the impact of this parameter and show the results in Figure \ref{fig:ImpactN}(b). The results show our attacks obtain better performance when $\gamma$ and $\beta$ are not too large (e.g., 1.0) or not too smaller (e.g., $<0.6$). One reason is too small values underestimate the effect of explanatory edges, while too large values overestimate the effect.  

\emph{Impact of $\xi$:} Figure \ref{fig:impactxi} shows the results with different perturbation budgets $\xi$. We can see our attacks become more effective with larger $\xi$. This is because a larger $\xi$ ensures the adversary has more capability to perturb the input graph.

\begin{table}[!t]
	\centering
\scriptsize	
 \addtolength{\tabcolsep}{-5.5pt}
	\caption{Attack performance of transferring the generated perturbed graphs on the perturbation-based GNN 
 explainer by baseline and our attacks 
 to other types of GNN explainers. } 
 {
	\begin{tabular}{cccccccc}
		\toprule 
		Explainer& Method &House&Community&Cycle&OGBN-P&MUTAG&REDDIT\\ 
		\midrule 

		\quad & Random& 11.25\%&22.83\%&18.04\%& 2.71\%&18.45\%&16.85\%\\
        CAM & Kill-hot&13.13\%&27.41\%&25.42\%& 2.82\%&21.50\%&19.25\%\\
		\quad & Loss-based& 25.33\%& 46.65\%& 42.00\%& 30.31\%& 33.50\%& 32.50\%\\
		\quad & Dedu.-based&{\bf 25.42}\%&{\bf 51.96\%}&{\bf 43.50\%}& {\bf 35.23}\%&{\bf 37.50\%}&{\bf 35.50\%}\\
		\midrule  
		\quad & Random&12.71\%&17.77\%&17.63\%& 18.83\%&17.70\%&13.25\%\\
        SA & Kill-hot&15.83\%&32.37\%&18.33\%& 14.31\%&26.50\%&19.50\%\\
        \quad & Loss-based&27.38\%& 46.96\%& 28.33\%& 32.00\%& 47.50\%& 44.50\%\\
		\quad & Dedu.-based&{\bf 41.25}\%&{\bf 50.18\%}&{\bf 34.17\%}& {\bf 41.54}\%&{\bf 53.50\%}&{\bf 55.25\%}\\
		\midrule 
		\quad & Random&11.45\% &18.48\%&11.67\%& 21.32\%&14.85\%&18.75\%\\
        GraphSVX&Kill-hot &  22.17\%&26.29\% &21.25\%&24.15\%&27.50\%&26.75\%\\
        \quad & Loss-based&31.46\%& 45.67\%& 23.52\%& 28.77\%&31.50\%& 33.06\%\\
		\quad & Dedu.-based&{\bf 35.42}\%&{\bf 49.69\%}&{\bf 26.46\%}&  {\bf 30.62} \%& {\bf 34.50\%}& {\bf 36.67\%}\\
		\midrule
		\quad & random&25.21\% &24.93\% &14.21\%&  5.88\% &11.40\% &15.45\%\\
  GEM & Kill-hot&29.58\% &29.64\% &19.17\%&  6.38\%&19.50\% &20.25\%\\
        \quad & Loss-based&{\bf 52.92\%}& 40.89\%& {\bf 38.50\%}& 28.50\%&36.50\%& 38.25\%\\
		\quad & Dedu.-based& { 47.71\%}& {\bf 44.91\%}&{ 36.50\%}&  {\bf 29.38} \%&{\bf 39.50\%}&{\bf 40.75\%}\\
        \bottomrule  
	\end{tabular}
 }
	\label{table:resultstransfer}
\vspace{-2mm}
\end{table}

\begin{table}[!t]
	\centering
 \small
 \renewcommand{\arraystretch}{0.99}
 \addtolength{\tabcolsep}{-4pt}
	\caption{Average runtime of our attacks.
 }
    \begin{scriptsize}
        
 {
	\begin{tabular}{ccccccc}
		\toprule 
		Method &House&Community&Cycle&OGBN-P&MUTAG&REDDIT\\ 
		\midrule  
		Loss-based&6.3s&39.4s&2.5s&124.1s&3.6s&1.8s\\
		Dedu.-based&7.7s&44.8s&2.7s&147.4s&4.0s&2.5s\\		
        \bottomrule  
	\end{tabular}
 }
    \end{scriptsize}
	\label{table:runtime}
\vspace{-2mm}
\end{table}

{\bf Transferability results:} Our attacks are 
designed mainly for perturbation-based GNN explainers. 
In this experiment, we study the effectiveness of the generated perturbed graphs by our attacks (default setting) on other four types of GNN explainers (See Section~\ref{supp:relatedwork} in Appendix). We choose decomposition-based CAM~\cite{pope2019explainability}, gradient-based SA~\cite{baldassarre2019explainability}, surrogate-based GraphSVX~\cite{DBLP:journals/corr/abs-2104-10482/GraphSVX}, and generation-based GEM~\cite{GEM} for evaluation. Table~\ref{table:resultstransfer} shows the results. 
As a comparison, we also show results on the two baselines.  
We can see the transferability results are not good with the 
random and kill-hot attacks.  
Instead, all explainers are more vulnerable to the perturbed graphs generated by our (deduction-based) attack.
Such results validate the promising transferability of our deduction-based attack is due to its produced attack edges are effective in general.

{\bf Runtime:} Theoretically, the worst-case complexity of our attacks against GNN explainers 
per graph is $O(T|V|^{2})$, where $T$ is the number of iterations to score the addition/deletion masks and $V$ is the number of nodes in the graph. This is because, in the worst case, our attacks need to find all addition/deletion candidates using a complete graph. 
Note that this complexity is the same as the GNN explainers themselves\footnote{To handle large graphs, the current GNN explainers only use the node’s local computation graph (i.e., a subgraph formed by the node’s within 2 or 3-hop neighboring nodes and their connections) for explanation. This can significantly scale down the graph size to be processed. Our attacks also perform on this subgraph.}. Empirically, Table~\ref{table:runtime} shows the average runtime on the 6 datasets in the default setting without any GPU (Macbook with a 2.30GHz CPU and an 8GB RAM).

\begin{table}[!t]
\renewcommand{\arraystretch}{0.95}
\centering
\vspace{-2mm}
\scriptsize
 \addtolength{\tabcolsep}{-2pt}
\caption{Black-box (deduction-based) attack with a surrogate loss.}
\begin{tabular}{@{}lcccccc@{}}
\toprule
           & House  & Community & Cycle   & OGB-P  & MUTAG  & REDDIT \\ \midrule
GNNExp.    & 10.83\% & 28.75\%   & 58.33\% & 25.42\% & 42.50\% & 45.53\% \\
PGExp.     & 33.75\% & 16.88\%   & 34.48\% & 23.89\% & 49.00\% & 42.63\% \\
GSAT       & 40.21\% & 30.49\%   & 37.36\% & 25.43\% & 47.50\% & 38.68\% \\
\bottomrule
\end{tabular}
\label{tbl:bbattack}
\vspace{-6mm}
\end{table}

\section{Discussion}

{\bf Black-box attack on GNN explainers:} 
We also test the black-box setting where the attacker is unknown to the explanation loss. To simulate this, the attacker uses a surrogate explanation loss, e.g., Cross Entropy loss + mask loss $\sum_{i \in E} |m_i| $).  
Table~\ref{tbl:bbattack} shows the results with our deduction-based attack using the surrogate loss. We can observe this attack still performs (much) better than the random and kill-hot baselines (see Table~\ref{table:result}), but is inferior to our loss-based and deduction-based attacks with known explanation loss.

{\bf Defending against the proposed attack:} 
We test two empirical defenses, but found they are not effective enough. 

\emph{Structure difference detection:} 
This defense aims to detect whether adversarially perturbed graphs generated by our 
attack are structurally different from 
clean graphs. Specifically, we utilize the structure similarity metrics NetSim and DeltaCon proposed in \cite{wills2020metrics}. We found all the perturbed graphs generated by our deduction-based attack and the respective clean graphs have a similarity larger than 0.95. This shows it is hard to differentiate perturbed graphs from clean ones, since  the generated perturbed  graphs are stealthy under the constraint in Equation (\ref{eqn:c3}).

\emph{Only API access:} This defense limits the attack knowledge by 
only providing the attacker the API access to target GNN explainers. 
However, it is still possible for the attacker to break this defense. 
For instance, the attacker can use a \emph{surrogate explanation loss} and perform our attack. 
Table~\ref{tbl:bbattack} shows this surrogate loss based attack is somewhat effective. 

We emphasize the designed attack uses limited knowledge about GNN explainers. Hence, a robust explainer that can defend against this \emph{weak} attack does not imply it is effective against stronger attacks. 
To verify whether a GNN explainer is intrinsically robust, it should test on the worst-case scenario, i.e., the white-box attack whether the attacker has full knowledge about the GNN explainer. A potential direction is hence to design provably robust GNN explainers.

\emph{Provably robust GNN explainers:} All the existing provable defenses for GNNs~\cite{bojchevski2020efficient,wang2021certified,zhang2021backdoor,jin2020certified,yang2023graphguard} focus on {classification models} that map an input graph to a \emph{scalar} label/confidence score. 
However, GNN explainers essentially map an input graph to a \emph{matrix} mask, making all  existing defenses inapplicable. 
We will leave designing provably robust GNN explainers as future work.

{\bf Attacking other types of GNN explainers.}
Different types of GNN explainers use different explanation mechanisms. To specifically attack these explainers, we need to redesign the attack objective and solution to fit each of them. Note that the transferability results in Table \ref{table:resultstransfer} have somewhat demonstrated the effectiveness of the adversarial edges generated by our attack on attacking all the other types of GNN explainers. 
It is valuable future work to study the vulnerability of other types of GNN explainers by designing the customized attack objectives and solutions.

\section{Conclusion}

We perform the first study on understanding the robustness of GNN explainers under graph structure perturbations, with an emphasis on perturbation-based GNN explainers. 
We formulate the attack under a realistic threat model, i.e., the attacker has limited knowledge about the explainer, allows small perturbation and maintains graph structure, and ensures correct GNN predictions. We propose two attack methods (loss-based and deduction-based) to realize the attack.  
Extensive evaluations validate existing GNN explainers are vulnerable to our proposed attacks. 
Future works include designing attacks against non-perturbation-based GNN explainers; and 
designing (provably) robust GNN explainers.

\section*{Acknowledgments}
We thank the anonymous reviewers for their insightful reviews and discussions. We thank Youqi Li for polishing the proof. 
Pang is supported in part by the Natural Science Foundation of Jiangxi Province of China (20232BAB212025) and High-level and Urgently Needed Overseas Talent Programs of Jiangxi Province (20232BCJ25024). 
Wang is supported in part by the Cisco Research Award and 
the National Science Foundation under grant Nos. ECCS-2216926, CCF-2331302, CNS-2241713 and CNS-2339686. 
Any opinions, findings and conclusions or recommendations expressed in this material are those of the author(s) and do not necessarily reflect the views of the funding agencies.

\section*{Impact Statement}

Graph Neural Networks (GNNs) become the mainstream methodology for learning graph data. 
Explainable GNNs aim to  foster the trust of using GNNs. However, this paper shows that state-of-the-art GNN explainers can be broken by practical, stealthy, and faithful adversarial attacks. 
Such vulnerabilities raise the concerns of using GNNs for safety/security-critical applications such as disease diagnosis and malware detection. 
Our findings urge the community to design more (provably) robust GNN explainers.

\bibliography{maindraft}
\bibliographystyle{icml2024}

\newpage
\appendix
\onecolumn

\section{Proof of the NP-Hardness of our Problem in Equations (\ref{problem})-(\ref{eqn:c4})}
\label{app:proveNP}

{Generally speaking, there are two approaches to prove the NP-hardness of a problem: 1) following the definition of NP-hardness, 2) reduction from a special case which is a well-known NP-hard problem \cite{gao2017opportunistic}. Here, we follow the latter one to prove the NP-hardness of our problem via the 0-1 knapsack problem, which is  NP-hard.}

For ease of description, we reformulate our attack problem via a matrix representation. We use the binary matrix ${\bf M}$ and $\tilde{{\bf M}}$ to denote the explanation result outputted by a parameterized GNN explainer (e.g., $g_\theta$) before and after the attack (i.e.,  ${\bf M}_{(i,j)}=1$ if the edge $(i,j)\in E_{S}$, otherwise ${\bf M}_{(i,j)}=0$; similarly, $\tilde{{\bf M}}_{(i,j)}=1$ if the edge $(i,j)\in \tilde{E}_{S}$, otherwise $\tilde{{\bf M}}_{(i,j)}=0$). We use ${\bf A}$ and $\tilde{{\bf A}}$ to represent the adjacency matrix of a graph $G=(V,E)$ and the perturbed counterpart $\tilde{G}=(V, \tilde{E})$, and a binary matrix ${\bf S} \in \{0,1\}^{|V|\times |V|}$ to indicate whether edges in ${\bf A}$ are perturbed or not (i.e. $\tilde{{\bf A}}={\bf A}\oplus{\bf S}$, $\oplus$ is the \texttt{XOR} operator). 
Based on these relation, we have ${\bf M} = g({\bf A})$ and $\tilde{\bf M} = g(\tilde{\bf A}) = g({\bf A}\oplus{\bf S})$. 

Finally, we use a matrix ${\bf W}$ to represent the cost for perturbing each edge: for $(i,j) \notin E_{S}$, ${\bf W}_{(i,j)}=1$; for $(i,j) \in E_{S}$, ${\bf W}_{(i,j)}\approx +\infty$, which is to satisfy the constraint in Eqn 9 of protecting explanatory edges from being modified. Then, our attack problem  can be rewritten as solving ${\bf S}$ as below: 
\begin{align}
&\underset{{\bf S} \in \{0,1\}^{|V|\times |V|}}{{\arg\min}} \, \sum\limits_{(i,j)\in |V|\times|V|} {\bf M}_{(i,j)}*\tilde{{\bf M}}_{(i,j)}
\label{problem2} 
\\
s.t.\quad 
& \sum\limits_{(i,j)\in |V|\times|V|} {\bf W}_{(i,j)}{\bf S}_{(i,j)}\leq \xi  \label{eqn:c2_2} \\
& \Lambda({\bf A},{\bf A}\oplus{\bf S}) < \tau  \label{eqn:c3_2} \\
& f({\bf A}\oplus{\bf S}) = y \label{eqn:c4_2}
\end{align}
This problem is reduced from
the 0-1 knapsack problem (\url{https://en.wikipedia.org/wiki/Knapsack_problem}), which is NP-hard. 
Next, we provide a more complete and detailed proof. 

Our key idea to prove our problem being NP-hard is to prove one of its subproblems to be NP-hard under certain requirements. 
Here, we 
generate a subproblem that is a 0-1 knapsack problem, where the variables ${\bf S}$ 
{\color{blue}} are
considered as "items" and the (bounded) "cost" is generated with our assumptions made on the explainer $g$, the GNN model, and the value of $\tau$ and $\xi$.   

First, we make an assumption on the explainer $g$. 
For every non-existent edge $e\in E_{A}$ ($= V \times V \setminus E$), we define a corresponding edgeset $\hat{E}_{e} \subseteq E \setminus E_{S}$ (and $\hat{E}_{e}$ can be a null set) and require: 
$$\sum\limits_{e\in E_{A}} |\hat{E}_{e}|\leq k$$
$$\forall e,e'\in E_{A}, e\neq e' \rightarrow \hat{E}_{e}\cap \hat{E}_{e'} = \emptyset$$

When explaining any perturbed graph $\tilde{G}=(V, \tilde{E})$, $g$ is assumed to follow the  rules:
(1) For any $e \in E_{A}$ added into $\tilde{E}$, $g$  selects all edges in $\hat{E}_{e}$ into explanation; (2) if the explanation edges are not full (i.e., less than $k$), $g$ chooses the remaining edges from $E_{S}$. 
With this, we define an matrix $\textbf{V}$ as below:
$$\textbf{V}_{(i,j)} = \left\{ 
    \begin{aligned}
    |\hat{E}_{(i,j)}|, &\quad (i,j) \in E_{A}\\
    0, &\quad \text{otherwise.}
    \end{aligned}
\right.$$
where $\textbf{V}_{(i,j)}$ represents the number of changed explanation edges by perturbing $(i,j)$ and its value ranges from 0 to $k$. 

Then, Eq (\ref{problem2}) could be rewritten as:
\begin{align}
\underset{{\bf S} \in \{0,1\}^{|V|\times |V|}}{{\arg\max}} \, \sum\limits_{(i,j)\in |V|\times|V|} {\bf V}_{(i,j)} * \textbf{S}_{(i,j)}
\end{align}

Next, we assume $f$ is a binary GNN classifier (parameterized by ${\bf W}$) and the prediction is based on: 
$$f({{\bf A}\oplus {\bf S}}) = \left\{ 
    \begin{aligned}
    &1, \quad \sum\limits_{(i,j)\in |V|\times |V|}\textbf{W}_{(i,j)} ({\textbf{A}}_{(i,j)} \oplus {\bf S})  \geq0\\
    &0, \quad \text{else}
    \end{aligned}
\right.$$
where $\textbf{W}$ 
assumes to have the same size as ${\bf A}$. 

We now define a constant matrix $\textbf{C}$ by: 
$$\textbf{C}_{(i,j)} = \left\{ 
    \begin{aligned}
    \textbf{W}_{(i,j)}, &\quad \text{if } \, \textbf{A}_{(i,j)}=1\\
    -\textbf{W}_{(i,j)}, &\quad \text{if } \, \textbf{A}_{(i,j)}=0
    \end{aligned}
\right.$$
Assume the prediction of the graph $G$ as $y=1$. Then, the constraint $f({\bf A}\oplus {\bf S})=y$ could be rewritten as:
$$\sum\limits_{(i,j)\in |V| \times |V|} \textbf{C}_{(i,j)}\textbf{S}_{(i,j)}\leq \sum\limits_{(i,j)\in |V| \times |V|}\textbf{W}_{(i,j)} {\textbf{A}}_{(i,j)}$$

By making further assumptions that $\tau = +\infty$ and $\xi = +\infty$, the problem is expressed as:
\begin{align}
&\underset{{\bf S} \in \{0,1\}^{|V|\times |V| \setminus E_S}}{{\arg\max}} \, \sum\limits_{(i,j)\in |V|\times|V|} {\bf V}_{(i,j)}\textbf{S}_{(i,j)}
\\
s.t.\quad 
& \sum\limits_{(i,j)\in |V| \times |V|} \textbf{C}_{(i,j)}\textbf{S}_{(i,j)}\leq \sum\limits_{(i,j)\in |V|\times |V|}\textbf{W}_{(i,j)} {\textbf{A}}_{(i,j)} 
\end{align}
In this formulation, $\textbf{V}_{(i,j)}\in \{0,\cdots, k\}$, $\textbf{C}_{(i,j)}\in (-\infty,+\infty)$ and $\textbf{S}_{i,j}\in \{0,1\}$ correspond to the value, weight, and item in the 0-1 knapsack problem, which   
cannot be solved in the polynomial time. 

{Note that the above reductions can be completed in finite steps, and the special case belongs to the NP-hard class. The complicated cases should be harder than the special case since more complex constraints are considered in our problems.} 
Therefore, our problem is ensured to be NP-hard.

\section{More Related Work}
\label{supp:relatedwork}

(i) \textit{Decomposition-based methods} consider the prediction of the model as a score and decompose it backward layer-by-layer until it reaches the input. The score of different parts of the input can be used to explain its importance to the prediction. 
Such methods include Excitation-BP~\cite{pope2019explainability}, DEGREE~\cite{feng2023degree}, CAM~\cite{pope2019explainability} and GNN-LRP~\cite{schnake2021higher}.

(ii) \textit{Gradient-based methods}, including SA~\cite{baldassarre2019explainability}, Guided-BP~\cite{baldassarre2019explainability} and Grad-CAM~\cite{pope2019explainability}, take the gradient of the prediction against the input to show the sensitivity of a prediction to the input. 
The sensitivity can be used to explain the input for that prediction.

(iii) \textit{Surrogate-based methods} replace the complex GNN model as a simple and interpretable surrogate model.
Representative methods are PGMExplainer~\cite{vu2020pgm}, GraphLime~\cite{huang2022graphlime}, GraphSVX~\cite{DBLP:journals/corr/abs-2104-10482/GraphSVX}, RelEx~\cite{zhang2021relex}, and DistilnExplain~\cite{pereira2023distill}.

(iv) \textit{Generation-based methods} 
use generative models or graph generators to generate explanations from the instance or model level. 
These methods include SubgraphX~\cite{DBLP:journals/corr/abs-2102-05152/subgraphX}, GEM~\cite{GEM}, RGExplainer~\cite{shan2021reinforcement/RGExplainer} and RCExplainer~\cite{Wang_2023/RCExplainer}. 
For instance, RCExplainer
applies reinforcement learning to generate subgraphs as explanations. In every step the agent takes an action to drop or add an edge from the current subgraph, and receives reward based on the new one. When it terminates, the final subgraph is used as the explanation.

(v) \textit{Perturbation-based methods
} aim to find the important subgraph or/and features as explanations by perturbing the input graph. Well-known methods include GNNExplainer~\cite{GNNEx19}, PGExplainer~\cite{DBLP:journals/corr/abs-2011-04573/PGExplainer}, GraphMask~\cite{schlichtkrull2021interpreting}, Zorro~\cite{funke2022z}, Refine~\cite{wang2021towards} and GStarX~\cite{zhang2022gstarx}.
In the paper, we mainly focus on {perturbation-based methods}: their explanation results are more accurate as they use both the original graph and target GNN model.

\begin{table*}[!t]
	\centering
	\caption{Loss, constraint, and mask initialization of representative perturbation-based GNN explainers.}
 \footnotesize
	\begin{tabular}{cccc}
		\toprule
		Explainer & Prediction Loss & Constraint Function & Learning Process\\ 
		\midrule
		GNNExplainer~\cite{GNNEx19} & Cross entropy &$\sum_{i\in E} |m_{i}|+m_{i} \log{m_{i}}+(1-m_{i}) \log{(1-m_{i})}$& Deterministic\\
		PGExplainer~\cite{DBLP:journals/corr/abs-2011-04573/PGExplainer} & Cross entropy &$\sum_{i\in E} |m_{i}|+m_{i} \log{m_{i}}+(1-m_{i}) \log{(1-m_{i})}$& Stochastic\\
        GSAT~\cite{DBLP:journals/corr/abs-2201-12987/GSAT} & Cross entropy &$\sum_{i\in E} m_{i} \log\frac{m_{i}}{r}+(1-m_{i}) \log\frac{1-m_{i}}{1-r}+c(n,r)$& Stochastic\\
        \bottomrule
	\end{tabular}
	\label{table:explainer}
\end{table*}

\section{More Experimental Settings}
\label{supp:ExperimentSetting}

\begin{figure}[!t]
  \centering
  \includegraphics[width=0.85\linewidth]{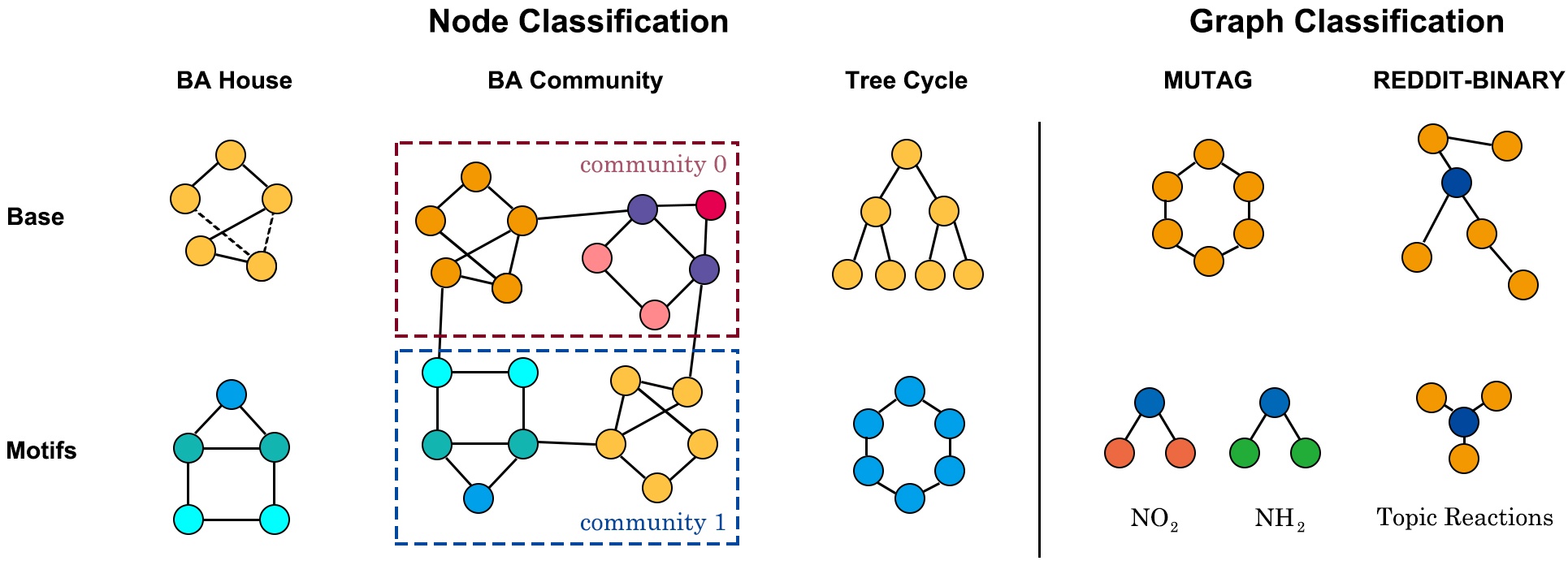}
  \vspace{-2mm}
  \caption{Example synthetic graphs and real-world graphs and their groudtruth explanations. 
  }
  \label{fig:data_des}
\end{figure}

{\bf Dataset description:} 
We provide more details of the datasets used in Section~\ref{sec:exp}. Some example graphs are in Figure~\ref{fig:data_des}.

\emph{BA House:} This graph stems from a base Barabási-Albert (BA) graph attached with ``house"-structured motifs as the groundtruth explanation. Nodes within the base graph have a label 0, while nodes positioned at the top, middle, or bottom of the ``house"-motif have labels 1, 2, and 3, respectively.

\emph{BA Community:} This dataset has two BA-Shapes graphs with two ``house"-motifs. 
Nodes are labelled based on their structural roles and community memberships, with eight classes in total.

\emph{Tree Cycle:} At its core lies a balanced binary tree. Six-node cycle motifs are appended to randomly chosen nodes from the base graph. The cycle motifs is the groundtruth explanation.

\emph{OGBN-Products:} This dataset is an undirected unweighted graph, representing an Amazon product co-purchasing network. Nodes represent products sold in Amazon, and edges between two products indicate that the products are purchased together.

\emph{MUTAG:} It consists of 4,337 molecular graphs, labeled in alignment with their mutagenic effect on the Gram-negative bacterium, Styphimurium. $NO_2$ and $NH_2$ motifs are considered as the groundtruth. 

\emph{Reddit-Binary:} It includes 2,000 graphs, each corresponding an online discussion thread on Reddit. Within a graph, nodes signify users, and edges denote interactions between them. The label of a graph is indicative of the interaction type within its thread. A star shape motif is considered as the groundtruth. 

\begin{table}[!t]
    \centering
    \vspace{-2mm}
    \caption{Parameter Setting. }
    \renewcommand{\arraystretch}{1}
    \footnotesize
\addtolength{\tabcolsep}{-5pt}
    \label{table:setting}
    \begin{tabular}{cccccccccc}
        \toprule
        Task& Dataset & \#Cases &$|E|_{\text{avg}}$&$k$&$\xi$&$N$&$\beta$ & $\gamma$\\
        \midrule
        \quad&BA House & 80 & 4110 & 6 & 5  & 4& 0.7 & 0.7 \\
       Node &BA Community & 160 & 8920 & 28 & 10  & 4 & 0.7 & 0.7\\
       Explanation &Tree Cycle & 160 & 1950 & 6 & 2  & 4& 0.7 & 0.7 \\
        \quad&OGBN-Prod. & 40 & 126036 & 25 & 10  & 4& 0.7 & 0.7 \\
        \midrule
        Graph &MUTAG & 40 & 97 & 5 & 2 &  4& 0.7 & 0.7 \\
        Explanation &Reddit-Bin. & 40 & 152 & 10 & 2 & 4& 0.7 & 0.7 \\
        \bottomrule
    \end{tabular}
  \vspace{-4mm}
\end{table}

{\bf Parameter setting:} Table~\ref{table:setting} shows the default parameter settings in our experiment.

\section{Algorithm Details}
\label{supp:loss-based}
\label{supp:deduction-based}

Algorithm~\ref{alg:loss} and Algorithm~\ref{alg:deduction} describe the detailed implementation of our two proposed attack algorithms.

\begin{algorithm}[!t]
\caption{Loss-based Attack}
\begin{small}
\begin{flushleft}
\label{alg:loss}
\textbf{Input}: A graph $G = (V,{E})$ with a label $y$, explanatory edges $E_S$ outputted by a GNN explainer, perturbation budget $\xi$, ratio test parameter $\tau=0.000157$, the explainer loss $\mathcal{L}$, and constant $\gamma$. \\
\rule{8.5cm}{0.7pt}
\end{flushleft}
\begin{algorithmic}[1] 
\STATE Initialize the maximum set difference $md = 0$
\STATE Initialize a mask $\tilde{M}^D$ for all edges in ${E}$ 
\STATE Initialize a filter matrix ${\bf f}^D$ and a bias matrix ${\bf b}^D$ for all edges in ${E}$: for $e \in E_S$, ${\bf f}^D_{e}=0$, ${\bf b}^D_{e}=1$; otherwise, 
${\bf f}^D_{e}=1$, ${\bf b}^D_{e}=0$
\STATE Set a loss function for $\tilde{M}^D \in [0,1]^{|E|}$ as $\mathcal{L}(\tilde{M}^D\otimes {\bf f}^D + \gamma \cdot {\bf b}^D)$. Learn $\tilde{M}^D$ by minimizing $\bar{\mathcal{L}}(\tilde{M}^D)$
\STATE Take $\xi$ edges from $E^D$ with the highest $\xi$ masked values in $\tilde{M}^D$ as deletion candidates $E_{del} = E^D.\text{top}_{\xi}(\tilde{M}^D)$
\STATE Initialize a mask $\tilde{M}^A$ for all edges in $E^A = E^C \cup E_S$
\STATE Initialize a filter matrix ${\bf f}^A$ and a bias matrix ${\bf b}^A$ for all edges in $E^A$: for $e \in E_S$, ${\bf f}^A_{e}=0$, ${\bf b}^A_{e}=1$; otherwise, 
${\bf f}^A_{e}=1$, ${\bf b}^A_{e}=0$ 
\STATE Set a loss function for $\tilde{M}^A$ as $\bar{\mathcal{L}}(\tilde{M}^A) = \mathcal{L}(\tilde{M}^A\otimes {\bf f}^A + \gamma \cdot {\bf b}^A)$. Learn $\tilde{M}^A$ by maximizing $\bar{\mathcal{L}}(\tilde{M}^A)$
\STATE Take $\xi$ edges from $E^A$ with the highest $\xi$ masked values in $\tilde{M}^A$ as addition candidates $E_{add} = E^A.\text{top}_{\xi}(\tilde{M}^A)$

\FOR{$0 \leq \xi_{A} \leq \xi$, $\xi_{D} = \xi - \xi_A$}
\STATE Add top $\xi_{A}$ edges in $E_{add}$ to ${E}$, resulting in $\tilde{E}_{add}$
\STATE Delete top $\xi_{D}$ edges in $E_{del}$ from $\tilde{E}_{add}$, resulting in $\tilde{E}$
\IF{$f(\tilde{G}) != y$ \OR $\Lambda(G,\tilde{G}(V,\tilde{E})) \ge \tau$}
\STATE {\bf Continue}
\ENDIF
\STATE Obtain new explanation $\tilde{E}_{S}$ on $\tilde{E}$ via Equations~\ref{ep} and \ref{exp_attack}
\IF{$|E_S - \tilde{E}_{S} \cap E_S | > md$ }
\STATE $md = |E_S - \tilde{E}_{S} \cap E_S|$
\ENDIF
\ENDFOR
\STATE {\bf return} $md$ and $\tilde{E}_{S}$
\end{algorithmic}
\end{small}
\end{algorithm}

\begin{algorithm}[!t]
\caption{Deduction-based Attack}
\begin{small}

\begin{flushleft}
\label{alg:deduction}
\textbf{Input}: A graph $G = (V,{E})$ with a label $y$, explanatory edges $E_S$ outputted by a GNN explainer, perturbation budget $\xi$, ratio test parameter $\tau$, the explainer loss $\mathcal{L}$, number of samples $N$, and $\beta$. \\
\rule{8.5cm}{0.7pt}
\end{flushleft}
\begin{algorithmic}[1] 
\STATE Initialize the maximum set difference $md = 0$
\STATE Initialize $\beta_{i}=\frac{i-1}{N-1}\times (1-\beta)+\beta$ for all $i \in \{1, 2, \cdots, N\}$
\STATE Initialize a mask $\tilde{M}^D$ for all edges in ${E}$
\STATE Initialize a filter matrix ${\bf f}^D$ and a bias matrix ${\bf b}^D$ for all edges in ${E}$: for $e \in E_S$, ${\bf f}^D_{e}=0$, ${\bf b}^D_{e}=1$; otherwise, 
${\bf f}^D_{e}=1$, ${\bf b}^D_{e}=0$
\STATE Set a loss function for $\tilde{M}^D \in [0,1]^{|E|}$ as $\bar{\mathcal{L}}(\tilde{M}^D) = \sum_{i=1}^N [ \mathcal{L}(\tilde{M}^D\otimes {\bf f}^D + \beta_{i} \cdot {\bf b}^D) - \mathcal{L}(\tilde{M}^D\otimes {\bf f}^D)]$. Learn $\tilde{M}^D$ by minimizing $\bar{\mathcal{L}}(\tilde{M}^D)$
\STATE Take $\xi$ edges from $E^D$ with the highest $\xi$ masked values in $\tilde{M}^D$ as deletion candidates $E_{del} = E^D.\text{top}_{\xi}(\tilde{M}^D)$
\STATE Initialize a mask $\tilde{M}^A$ for all edges in $E^A = E^C \cup E_S$
\STATE Initialize a filter matrix ${\bf f}^A$ and a bias matrix ${\bf b}^A$ for all edges in $E^A$: for $e \in E_S$, ${\bf f}^A_{e}=0$, ${\bf b}^A_{e}=1$; otherwise, 
${\bf f}^A_{e}=1$, ${\bf b}^A_{e}=0$ 
\STATE Set a loss function for $\tilde{M}^A$ as $\bar{\mathcal{L}}(\tilde{M}^A) = \sum_{i=1}^N [ \mathcal{L}(\tilde{M}^A\otimes {\bf f}^A + \beta_{i} \cdot {\bf b}^A) - \mathcal{L}(\tilde{M}^A \otimes {\bf f}^A)]$. Learn $\tilde{M}^A$ by maximizing $\bar{\mathcal{L}}(\tilde{M}^A)$
\STATE Take $\xi$ edges from $E^A$ with the highest $\xi$ masked values in $\tilde{M}^A$ as addition candidates $E_{add} = E^A.\text{top}_{\xi}(\tilde{M}^A)$

\FOR{$0 \leq \xi_{A} \leq \xi$, $\xi_{D} = \xi - \xi_A$}
\STATE Add top $\xi_{A}$ edges in $E_{add}$ to ${E}$, resulting in $\tilde{E}_{add}$
\STATE Delete top $\xi_{D}$ edges in $E_{del}$ from $\tilde{E}_{add}$, resulting in $\tilde{E}$
\IF{$f(\tilde{G}) != y$ \OR $\Lambda(G,\tilde{G}(V,\tilde{E})) \ge \tau$}
\STATE {\bf Continue}
\ENDIF
\STATE Obtain new explanation $\tilde{E}_{S}$ on $\tilde{E}$ via Equations~\ref{ep} and \ref{exp_attack}
\IF{$|E_S - \tilde{E}_{S} \cap E_S | > md$ }
\STATE $md = |E_S - \tilde{E}_{S} \cap E_S |$
\ENDIF
\ENDFOR
\STATE {\bf return} $md$ and $\tilde{E}_{S}$
\end{algorithmic}
\end{small}
\end{algorithm}

\begin{table}[!t]
\centering
\footnotesize
\caption{Average confidence difference of the most-likely class before and after our attack.}
\begin{tabular}{@{}lcccccc@{}}
\toprule
            & House & Community & Cycle & OGB-P & MUTAG & REDDIT \\ \midrule
GNNExp.     & 0.0015 & 0.0575  & -0.0495 & 0.0135 & 0.0005  & -0.0052 \\
PGExp.      & 0.0165 & 0.0148  & -0.0426 & 0.0156 & 0.0036  & 0.0039  \\
GSAT        & 0.0235 & 0.0160  & -0.0007 & 0.0145 & -0.0098 & 0.0051  \\
\bottomrule
\end{tabular}
\label{tab:confdiff}
\end{table}

\begin{table}[ht]
\centering
\footnotesize
\caption{Average number of different explanations edge outputted by GNN explainers in five runs.}
\begin{tabular}{@{}lcccccc@{}}
\toprule
            & House  & Community & Cycle & OGB-P & MUTAG & REDDIT \\ \midrule
GNNExp.       & 0.17/6 & 3.38/28   & 0.75/6 & 1.77/25 & 0.12/5 & 1.82/10 \\
PGExp.        & 0/6    & 0/28      & 0/6    & 0/25    & 0/5    & 0/10 \\
GSAT        & 0/6    & 0/28      & 0/6    & 0/25    & 0/5    & 0/10 \\
\bottomrule
\end{tabular}
\label{tab:consistency}
\end{table}

\section{More Discussions}
\label{supp:discussion}

{\bf Differences between attacking GNN classifiers and attacking GNN explainers:}
Both attacks on GNN classifiers and GNN explainers perturb the graph structure, but their attack goals (and hence attack objectives) are different. 
The former attack focuses on altering \emph{GNN predictions}, and the latter attack altering \emph{GNN explanations} while retaining \emph{GNN predictions}. 
Specifically,  
the former attack  perturbs the graph structure such that a target GNN classifier produces a wrong (node/graph) \emph{label prediction} on the perturbed graph. 
In contrast,  the latter attack perturbs the graph structure such that a target GNN explainer generates a wrong \emph{explanatory subgraph},  
while maintaining the correct GNN prediction on the perturbed graph. 

From the formulation of attack objective, the former attack optimizes the \emph{differentiable} loss used by a GNN classifiers, while the later optimizes the \emph{non-differentiable} edge set outputted by a GNN explainer, under certain constraints (this problem is proved to be NP-hard). 
A possible solution is to adapt the idea of existing attacks by relaxing our attack objective with an approximate differentiable loss (i.e., our loss-based attack), but our results show the attack performance is not promising. We then design a more accurate deduction-based attack that directly mimics the learning dynamics of the GNN explainer.

{\bf Do the proposed attacks affect the GNN prediction confidence?} We record the average confidence score difference of the most likely class before and after our deduction-based attack. Table~\ref{tab:confdiff} shows the results, where  positive/negative value means a decrease/increase. We see our attack  nearly has no influence on the score of the most confident class.

{\bf How consistent are the GNN explanation results, or whether the proposed attack is  statistically meaningful?}     
We tested the explanation consistency on the studied perturbation-based GNN explainers. For every (node/graph) instance, we first explain it once with a target GNN explainer to obtain $E_s$, and then rerun five times of the explainer. We calculate the average number of different edges (denoted as A) in these explanations compared with edges in $E_s$. The results of $A/|E_s|$ are shown in Table \ref{tab:consistency}. We observe GNNExp. has marginal inconsistency, while PGExp. and GSAT do not have the inconsistency issue. Hence, our attack results are statistically meaningful.

\end{document}